\documentclass[preprint]{aastex}
\shorttitle{Vibrationally-Excited Molecular Hydrogen in Cooling Flows}
\shortauthors{Donahue et al.}
\newcommand{\lta}{{\>\rlap{\raise2pt\hbox{$<$}}\lower3pt\hbox{$\sim$}\>}}
\newcommand{\gta}{{\>\rlap{\raise2pt\hbox{$>$}}\lower3pt\hbox{$\sim$}\>}}

\newcommand{\flux}{{~\rm erg \, s^{-1} \, cm^{-2}}}
\newcommand{\sbr}{{~\rm erg \, s^{-1} \, cm^{-2} \, arcsec^{-2}}}

\newcommand{\etal}{{et al.~}}
\newcommand{\lum}{{~\rm erg \, s^{-1}} }

\newcommand{\Mdot}{$\dot M$}

\begin{document}

\title {HST Observations of Vibrationally-Excited Molecular Hydrogen in Cluster Cooling Flow Nebulae 
\footnote{Based on observations with the NASA/ESA Hubble Space Telescope, obtained at the 
Space Telescope Science Institute, which is operated by the Association of Universities 
for Research in Astronomy, Inc.\ (AURA), under NASA contract NAS5-26555}}

\author{Megan Donahue, Jennifer Mack, G. Mark Voit, William Sparks}
\affil{Space Telescope Science Institute, 3700 San Martin Dr., Baltimore, MD 21218}
\email{donahue@stsci.edu,mack@stsci.edu,voit@stsci.edu,sparks@stsci.edu}
\author{Richard Elston}
\affil{University of Florida,Dept. of Astronomy, 211 Space Sciences Bldg., 
Gainesville, FL 32611-2055}
\email{elston@astro.ufl.edu}
\and
\author{Philip R. Maloney}
\affil{University of Colorado, CASA, CB-389, Boulder, CO 80309}
\email{maloney@casa.colorado.edu}

\begin{abstract}
We report the results of Hubble Space Telescope 
NICMOS and WFPC2 imaging of emission-line
nebulae in the central galaxies of three 
clusters of galaxies purported to host 
massive cooling flows, Perseus (NGC1275), Abell~2597, and PKS0745-191. 
The spectral signature of vibrationally-excited molecular hydrogen 
has been seen in every galaxy searched thus far that is 
central to a cluster cooling flow with an optical emission line nebula.
With the exquisite spatial resolution available to us with the Hubble Space
Telescope, we have discovered that the vibrationally-excited molecular hydrogen 
gas extends several kpc from the centers of Abell~2597 and 
PKS0745-191, while the vibrationally-excited molecular hydrogen in NGC1275 appears to
be mostly confined to its nucleus, with some extended emission 
$<1$ kpc from the center. The molecular hydrogen in Abell~2597 and
PKS0745-191 seems to be nearly co-spatial with the optical emission-line
filaments in those systems. 
There may be a tiny jet visible in the 1.6 $\mu$m image of PKS0745-191.
We also find significant dust absorption
features in the 1.6 $\mu$m images of all three systems. The dust lanes are
not strictly 
co-spatial with the emission-line filaments, but are aligned with and perhaps
intermingled with them. 
The morphology of the emission-line systems suggests that the presence
of vibrationally-excited molecular hydrogen is not purely an AGN-related property of 
cluster ``cooling-flow'' nebulae, and that the optical and infrared emission-line gas, 
that is, the ionized and vibrationally-excited molecular gas 
have similar origins, if not also similar energy sources.  
The infrared molecular hydrogen lines are much too bright to be generated
by gas simply cooling from a cooling flow; furthermore, 
the gas, because it is dusty, 
likely did not condense from the hot intracluster medium (ICM). We examine some 
candidates for heating the nebulae, including X-ray irradiation by the ICM, 
UV fluorescence by young stars, and shocks. 
UV heating by young stars provides the most
satisfactory explanation for the H$_2$ emission in A2597; X-ray irradiation
is energetically unlikely and strong shocks ($v\gta40$ km s$^{-1}$) are 
ruled out by the high H$_2$/H$\alpha$ ratios. 
If UV heating is the main energy input, a few billion solar masses 
of molecular gas is present in A2597 and PKS0745-191. UV irradiation models predict a
significant amount of $1.0-2.0$ micron emission line from higher excitation
H$_2$ transitions and moderate far infrared luminosities ($\sim10^{44} h^{-2}
\lum$) for A2597 and PKS0745-191. Even in the context of UV
fluorescence models, the total amount of H$_2$ gas and star formation inferred
from these observations is too small to account for the cooling flow rates 
and longevities inferred from X-ray observations. We note an interesting
new constraint on cooling flow models: the radio sources
do not provide a significant amount of shock heating, and therefore
they cannot counterbalance the cooling of the X-ray gas in the cores of these 
clusters.
 
\end{abstract}

\section{Introduction}

The high density of the hot intracluster medium (ICM) at the centers of 
many clusters of galaxies implies that this central ICM may cool within a 
Hubble time. The short cooling time of this gas  
prompted suggestions that the ICM in these clusters cools and flows 
towards the center, where it subsequently condenses (see Fabian 1994 and
references therein for a review.) The central galaxies of clusters
suspected to harbor such ``cooling flows'' frequently display luminous, 
extended emission line nebulae (e.g., Heckman \etal  1989), which appear to be
ionized and at least partially heated 
by hot, young stars (Voit \& Donahue 1997; Cardiel, Gorgas,
Aragon-Salamanca 1995, 1998; Hansen \etal  1995; Johnstone, Fabian \& 
Nulsen 1987). 
The star formation rates inferred from such observations are
similar to those in some starburst galaxies (e.g. McNamara 1997). 
The optical emission-line luminosities
from the ``cooling-flow'' nebulae  are strongly correlated (99.94\% confidence) 
with the mass flow rate inferred from the X-ray measurements, 
albeit with a dispersion spanning
up to two orders of magnitude (Heckman \etal  1989). The emission-line
nebulae are common, appearing in $\sim40\%$ of X-ray selected clusters
of galaxies (Donahue, Stocke \& Gioia 1992). The emission-line
nebulae in clusters seem to avoid clusters with central cooling times
longer than about a Hubble time (Hu \etal  1990). Yet the
optical emission line gas itself appears to be dusty (Donahue \& Voit 1993;
Sparks, Ford \& Kinney 1993; Sparks, Macchetto, \& Golombek 1989),
and thus is not likely to be a direct condensate from the ICM.

The presence of vibrationally-excited molecular hydrogen may be as prevalent as 
optical emission line nebulae in the central
galaxies of cluster cooling flows. All central cluster galaxies with large inferred mass 
deposition rates and powerful optical emission-line systems  
observed to date with sufficient sensitivity emit 
strong molecular hydrogen emission in the 2 micron 1-0 S(1) line 
(Elston \& Maloney, 1992; Elston \& Maloney 1994). More recent 
IR spectroscopic observations 
of radio galaxies associated with strong cooling flows reveal that these
galaxies produce H$_2$ emission 
while other radio galaxies do not (Falcke \etal  1998; Jaffe \etal  1997).  
The infrared spectra show H$_2$ 
vibrational line ratios characteristic of collisionally excited 
$\sim1000-2000$ K molecular gas.

Dusty nebular filaments do not only appear in cluster cooling flows. 
Such structures have also been detected in early type galaxies in small 
groups (e.g. Goudfrooij \& Trinchieri 1998; Singh \etal  1994), and in interacting
gas-rich galaxies (Kenney \etal  1995; Donahue \etal  in preparation). The
characteristics in common with cooling flow optical emission-line 
filaments are the size scales (kiloparsecs) and the prominent forbidden
line and recombination emission ([NII], H$\alpha$). Some of these sources
have had recent interactions. Most of these sources also have hot ISM or
are embedded in a cluster or group. In none of these objects is the source
of the filament energy understood.

This paper is organized as follows. 
Our targets are described in Section~\ref{sample}. 
In Section~\ref{observations} we describe the observations.  
Section~\ref{reduction} outlines the data reduction and analysis 
techniques for both
Hubble Space Telescope (HST) 
NICMOS and WFPC2 data including construction methods for isophotal
profiles, for continuum subtraction from the emission-line flux, and for
absolute flux calibration. In Section~\ref{aperture}, we report and
discuss the emission line fluxes inside the central 2" of each source.
In Section~\ref{maps}, 
we
compare the morphology of the molecular gas to the ionized gas, the 1.6 micron dust, and the
radio emission. In Section~\ref{dust}, we investigate the opacity and
reddening of dust at multiple wavelengths. We discuss the implications
of our results for the source in \S\ref{source} and for the heating
mechanism in \S\ref{heat}. We review our conclusions in \S\ref{conclusions}.
We also report our 1.6-$\mu$m photometry for the young clusters in 
NGC1275 in Appendix A.
Luminosities and angular distances are calculated  
assuming $q_0=0.5$ and $H_0 = 100h$ km s$^{-1}$ Mpc$^{-1}$.

\section{Sample Selection \label{sample}}

Three cluster targets, Perseus, Abell 2597, and PKS0745-191, were chosen for 
imaging with the Hubble Space Telescope (HST) 
NICMOS camera in order to study the relationship between the 
vibrationally-excited molecular hydrogen, the optical line-emitting gas, the dust, and
the radio source in the central few arcseconds of X-ray luminous clusters. 
NGC1275 is in the 
center of the Perseus cluster, and PKS0745-191 is the name of the radio
source in the cluster of the same name, but as a convention, 
we will refer to the central galaxy in Abell 2597 by the name of its 
host cluster, and not by the name of its radio source (PKS 2322-122). 
All three targets are embedded in cooling flows with
significant cooling rates. The nuclei of all three targets were known 
a priori to emit bright infrared H$_2$ lines whose observed
wavelengths fall within the bandpasses of narrow-band NICMOS filters. 
Star formation rates inferred from the blue optical excess 
in all 3 systems span $10-20 h^{-2}$ M$_\odot$ yr$^{-1}$ 
(McNamara, 1997).  Global target properties are summarized in
Table~\ref{targets}.

\begin{table}
\caption{Target Properties \label{targets}}
\begin{tabular}{lccccc}
Target     & Redshift & $\dot{M}h^{-1}$       & Scale & kT & $N_H$ \\
  & &  ($M_\odot$ yr$^{-1}$) & ($h^{-1}$ kpc/arcsec) & (keV) 
& $(10^{20} ~\rm{cm}^{-2})$     \\ 
\tableline \tableline
NGC1275    & 0.0175   &  250            &  0.24  & 6.3 & 14.5 \\
Abell2597  & 0.0852   &  130            &  1.07  & 8.5 & 2.5  \\
PKS0745-19 & 0.1028   &  450            &  1.26  & 9.1 & 46.1 \\ \tableline
\end{tabular}
\end{table}

NGC1275 is the nearest massive cooling flow, with an inferred cooling rate of 
$\approx200h^{-1}$ M$_\odot$ yr$^{-1}$ (Edge \& Stewart, 1991). It is rather atypical 
for a cooling  flow, however, with an luminous 
AGN in its center and a foreground, high-velocity system, possibly
a colliding galaxy (Hu \etal  1983). Because NGC1275 is known  
to be a notoriously complex albeit bright, cluster cooling flow
system, we also observed Abell 2597 and PKS0745-191, two somewhat  
more typical cooling flow cluster nebulae that 
have luminous, unresolved  H$_2$ emission. These two cluster galaxies 
have  among the most luminous 1-0 S(1) nuclear emission observed by Elston \& 
Maloney (1994).

In NGC1275, there is a large amount of cold molecular
gas -- $\sim 6\times10^{9}  \, {\rm M}_\odot$, assuming the debatable standard
CO/H$_2$ conversion -- most recently mapped in CO by 
Bridges \& Irwin 1998 and Braine \etal  1995. The CO emission 
was discovered by Lazareff \etal  1989. In contrast, 
CO from cold molecular gas has not been detected 
in PKS0745-19 (O'Dea \etal
1994), with an upper limit of $8\times10^{9} \, {\rm M}_\odot$, again assuming 
the standard conversion. McNamara \& Jaffe (1994) searched six other cooling flow clusters with
comparable mass cooling rates for a trace of cold molecular gas (CO), 
and found nothing with similar limits. Abell 2597 has not been
observed in the radio for CO, but notably, it is the only known cluster 
radio source other than NGC1275 with H~I absorption (O'Dea, Baum \&
Gallimore 1994).

Specific
comments regarding each source follow here.

\textbf{NGC~1275}: 
\objectname{NGC 1275} is a complex galaxy system with $z=0.01756$ (Strauss \etal  1992) 
(1 arcsec = 0.24 kpc h$^{-1}$, $4\pi d_L^2 = 3.34 \times10^{53} 
h^{-2}~\rm{cm}^2$). 
It is a Seyfert galaxy (Seyfert 1943), an IRAS galaxy (Strauss \etal  1992), a
Markarian galaxy, a blazar, the cD galaxy of the Perseus cluster (\objectname{Abell 426}), 
a merger system, and a radio source (\objectname{3C84} or \objectname{Perseus A}). 
It has a giant nebula of luminous 
optical emission-line filaments (cf. Burbidge \& Burbidge 1965)
extending $\sim100$ kpc from its center. \objectname{Perseus} is the closest example of a cluster
of galaxies with a massive cooling flow  with \Mdot $\sim 250 h^{-1}~{\rm M_\odot 
\, yr^{-1}}$ (Peres \etal 1998; Mushotzky \etal  1981; Edge \& Stewart 1991). Fischer \etal  (1987)
first discovered vibrationally excited molecular hydrogen in its nucleus.

\textbf{A2597}: 
The central galaxy of \objectname{Abell 2597} ($z=0.08520$, Struble \& Rood, 1987),
 ($4 \pi d_L^2 = 8.11\times10^{54} h^{-2}~\rm{cm}^{2}$,
1 arcsec = 1.07 $h^{-1}$ kpc),
also known as \objectname{PKS2322-123}, is in the center
of a massive cooling flow ($\dot{M} \sim  130 h^{-1}~{\rm M}_\odot$ yr$^{-1}$ 
(Crawford \etal  1989). 
It has a tiny double sided radio 
source (Sarazin \etal  1995) and small blue 
lobes (McNamara \& O'Connell 1993),
which are relatively unpolarized and thus are thought to be scattered
light or synchrotron emission (McNamara \etal  1999).
It has an extensive and 
luminous 
emission line nebula (Crawford \etal  1989; Heckman \etal  1989), 
whose emission lines are inconsistent
with being produced by shocks (Voit \& Donahue 1997).

\textbf{PKS0745-191}: 
\objectname{PKS0745-191} ($z=0.1028$, Hunstead, Murdoch \& Shobbrook 1978);  
($4 \pi d_L^2 = 1.19\times10^{55} h^{-2}~\rm{cm}^{2}$, 
1 arcsec = 1.26 $h^{-1}$ kpc) contains a strong
cooling flow, $\sim450 h^{-1}$~M$_\odot$ yr$^{-1} $ (Allen 2000;  Peres \etal 1998; Edge \& Stewart
1991; Fabian \etal 1985),
a powerful H$\alpha$ source (Fabian \etal 1985), and like Perseus and A2597, excess blue
light in the interior few kpc (McNamara \& O'Connell 1992; Fabian \etal 1985). 
It is radio-loud compared to NGC1275 $(L_R= 4.5 \times 10^{42} h^{-2}\lum$), 
with 5-10 times the radio luminosity of NGC1275 at 1.4 GHz (Baum \& O'Dea 
1991.)


\section{HST Observations \label{observations}}

High signal-to-noise observations of the molecular hydrogen 
and continuum were acquired with the 
HST NICMOS2 camera, with a pixel size of 0.075" and a field of
view of $19.2" \times 19.2"$, 
(Thompson \etal  1998) using narrow and broad-band filters. 
For NGC1275, the narrow band F216N filter allows the 
observation of the H$_2$ 1-0 S(1) line (rest wavelength 
2.12 microns). The other two targets were observed in the  
1-0 S(3) line  (rest wavelength 1.956 $\mu$m) 
whose nuclear flux is nearly equal to the 1-0 S(1) flux in these systems 
(Falcke \etal  1998; Elston \& Maloney 1994).
This line redshifts into the bandpasses of the F215N and F212N NICMOS filters
 for  PKS0745-191 and A2597 respectively. 
We obtained broad band F160W images for continuum measurements. 
The bandpass of the F160W filter ($\sim$H-band) 
is sufficiently wide and the emission lines 
sufficiently weak such that the F160W image is of nearly pure continuum light.

To image the optical line emission at a  
spatial resolution similar to our molecular-line maps, 
we acquired HST Wide-Field Planetary Camera 2 (WFPC2) 
data (Holtzman \etal 1995), both through our own program 
and from the Hubble Data Archive. The WF cameras have 0.0966" pixels
and $80" \times 80"$ field of view, while the PC camera has 0.0455" pixels 
and a $36" \times 36"$ field of view. 
For NGC1275, we retrieved archival WFPC2 data taken with the linear ramp 
filter (LRF) at 6676\AA\  
centered on H$\alpha$ and [N~II] of the low-velocity system. 
The LRF has a very narrow 
bandwidth and is ``tuned''  to the desired central wavelength 
by appropriate target placement in
the field of view.  For A2597 we 
used archival WFPC2 imaging of [OII]3727 emission through the F410M 
filter and 
of H$\alpha$+[NII] emission measured through the F702W filter. The 
stellar continuum  
images were taken through the F702W filter for NGC1275 and 
the F160W filter for A2597. We observed  
PKS0745-191 through the LRF 
for an H$\alpha$ image. 
Blue continuum imaging was carried out 
using the F439W filter.  
The total exposure times and observational
parameters for each observation are presented in Table~\ref{obslog}.

\begin{table*}
\caption{Observing Log\label{obslog}}
{\begin{tabular}{ll|lllr}
\tableline\tableline\\
\multicolumn{1}{l}{Target} &
\multicolumn{1}{l}{Camera} &
\multicolumn{1}{l}{Filter} &
\multicolumn{1}{l}{Date} &
\multicolumn{1}{l}{Datasets} &
\multicolumn{1}{l}{Exp. time (sec.)} \\
\tableline\\
NGC1275 &  NICMOS2  &  F216N  & 08/15/97  &N46001010 		&1536\\
	&  	    &  F160W  & 08/15/97  &N46001020  		&256\\
	&           &  F160W  & 03/16/98  &N3ZB1R010		&640\\
\\
	& WFPC2	    &  FR680N &	09/10/95  &U2S01601T 		&600\\
	&  	    &  F702W  &	03/31/94  &U27L1G01T-G04T       &560\\
	&           &  F702W  & 11/16/95  &U2PF0402T-405T	&4500\\
	&	    &  F450W  & 11/16/95  & U2P0407T-40AT	&4900 \\

\tableline\\
ABELL2597& NICMOS2  &  F212N  & 10/19/97  &N46003010-3020  	&12032\\
	&           &         & 12/03/97  &N46004010-4030  	&\\
	&  	    &  F160W  & 12/03/97  &N46004ZBQ 		&384\\
\\
	&  WFPC2    &  F410M  & 07/27/96  &U3CU0101T-104T       &2200\\
	&  	    &  F450W  & 07/05/95  &U2PF0203T-204T       &2500\\
	&  	    &  F702W  & 07/05/95  &U2PF0201T-202T       &2100\\

\tableline\\
PKS0745-191&NICMOS2 &  F215N  & 10/10/97  &N46002010-2030  	&7168 (3566)\tablenotemark{a}\\
	&	    &         &	09/06/98  &N46005010  		&\\
	&  	    &  F160W  & 10/10/97  &N46002040  		&512\\
\\
	& WFPC2	    &  FR680N &	10/11/97  &U460A201R-204R       &3200\\
	& 	    &  F439W  &	10/11/97  &U460A205R,206R	&1000\\
\tableline\\
\end{tabular}}

\tablenotetext{a}{Effective exposure time for the degraded NICMOS observation in parentheses.}

\end{table*}

\section{HST Data Reduction and Calibration \label{reduction}}
	\subsection{Image Reduction}

HST NICMOS {\em multiaccum} 
data may present the observer with significant data reduction challenges, 
which could not 
be handled within the current standard HST pipeline software called {\em calnica}. 
In the following section we describe our infrared data reduction 
and calibration process, using tasks in the Space Telescope 
Science Data Analysis System 
(\anchor{http://ra.stsci.edu/STSDAS.html}{STSDAS}, v2.1.1), obtained 
as a standard package in the publically-available 
Image Reduction and Analysis Facility 
(\anchor{http://iraf.noao.edu/iraf-homepage.html}{IRAF}, v2.11.3). 
The solutions we found may prove a useful guide 
to readers reducing HST NICMOS data, 
therefore we describe our reductions in moderate detail, citing the
software tasks whenever possible. We assume a 
convention of naming packages in capitals and tasks in italics. We also 
estimate the magnitude of the systematic uncertainties arising from imperfect 
data reduction and calibration.

We calibrated the raw NICMOS data and corrected it for various instrument artifacts
following methods which were in development by the NICMOS instrument group at STScI. 
Images were processed
through the first portion of the STSDAS program {\em calnica}, producing images 
which are corrected for dark and bias signals and for nonlinearity. Some  
additive noise, the dark current ``pedestal'', still exists after this correction and 
must be removed before the flatfielding step 
to prevent imprinting an inverted response pattern on 
the final image. 

To quantify the zeroth-order pedestal in each quadrant we estimated 
the mean background signal, excluding sources, 
which consists of the sky signal plus a constant, quadrant-dependent bias. 
The initial estimate of the bias level was multiplied by the flatfield and 
subtracted. This process was done 
independently for each readout frame and quadrant of 
the total NICMOS exposure. The median residual was then subtracted from each 
quadrant, and the root mean square variation of the resulting 
background was measured. This process was repeated for 
a range of initial guesses to 
deduce the value which minimized the root mean square variation 
of the background and thus, of 
the influence of the large-scale 
flatfield pattern. It was necessary to be very careful when subtracting 
the background from these images, because the outer isophotes 
of the galaxy also contribute 
to the large-scale background signal.
When the source fills the majority of the frame, separating the bias, sky background,
and galaxy light is more difficult. We obtained our best results when we
subtracted the pedestal after  
removing the target galaxy by fitting its surface brightness profile and 
subtracting the profile from the data.  

A residual, noiseless spatial variation, called ``shading'',  in the  
bias after dark subtraction remained in only a few of the 
narrow-band A2597 images, resulting in a position-dependent, low-level 
background. The bias changes slowly 
across the field of view, and again, if not removed, it  
results in an imprint
of the flat-field pattern left in the data. 
The ``shading'' pattern results because the reference dark 
image does not precisely reproduce the temperature-dependent 
dark current if the temperature of
the detector varied or was different from that of the reference dark image.
 
To correct for this small effect, we modelled 
the background outside the central 7.5" by 7.5" box by median filtering the image with a boxcar size of $13\times13$ pixels, 
and subtracted the resulting median background 
from the pixels outside the central box region. 
This procedure, mostly cosmetic, brings out the contrast of the 
central structures without changing the value of the mean background.
The effect of not fully correcting for this additional 
additive background in the central 7.5" adds an 
additional systematic uncertainty in the absolute H$_2$ emission-line 
fluxes for A2597 of $\sim 5\%$ after flat-fielding. 
(Systematic uncertainties in the H$_2$ fluxes 
arising from calibration are $\lesssim5\%$ without this effect.)

After we estimated and subtracted the effective dark current, the standard 
{\em calnica} process 
flatfielded the data, corrected for cosmic
rays, and converted the units to counts (or DN) s$^{-1}$. 

The final challenge was to correct for cosmic ray persistence.  NICMOS images 
taken very soon after HST passes through the South Atlantic Anomaly  
radiation
belt are affected significantly by 
cosmic ray persistence and thus are degraded by noise. The  
decaying afterglow from energetic cosmic rays persist from one read of the 
detector to the next, causing the signal from a single cosmic ray 
event to persist from one read to the next. 
Two-thirds of our narrow band images of PKS0745 were significantly impacted by persistence.  We took significant amounts of off-source
data for the narrow-band observations, originally to enable the
removal of any significant thermal
component at 2 microns. 
To remove persistant cosmic ray contamination, we modelled the charge 
decay rate by scaling the off-source images taken later in time and subtracting the scaled
image from the 
initial on-source image. This procedure 
was iterated until we found 
a constant value which minimized the root mean square 
variation of the background. 
For our affected images, this procedure improved 
the overall signal to noise by about 30\%. Finally, we aligned these processed images and
combined them into a single weighted average image.

WFPC2 data are much easier to calibrate, and the standard STSDAS 
pipeline procedure {\em calwp2}, 
with the standard updated reference files were used to 
process our data. The linear-ramp-filter (LRF) images were flatfielded with
the narrow band filter flat of the nearest available 
wavelength (filter F673N) (\anchor{http://www.stsci.edu/instruments/wfpc2/Wfpc2_isr/wfpc2_isr9606.html}
{WFPC2 Instrument Science Report 96-06}).
Images were combined and cosmic rays removed using the STSDAS task {\em crrej}.

	\subsection{Isophotal Fitting and Continuum Subtraction \label{isofit}}

Isophotal fitting of the NICMOS data was performed for the inner 7" radial
region of each galaxy using 
the STSDAS task {\em ellipse}. All parameters (ellipticity, position angle, 
and center position) were left free to fit the H-band (F160W) 
continuum. A continuum model was created using the best-fit parameters. A similar
ellipsoidal fit was made to the narrow-band data, but with the ellipticity, position 
angle
and centroid fixed to those parameters modelling the continuum data.  
The difference between the narrow band data and the continuum 
model, appropriately scaled,
provides an estimate of the total amount and extent of the residual H$_2$ emission. The relative
scale factor (Table~\ref{cal_dat}, Isophotal Counts Ratio) 
was determined by matching the galaxies' isophotal profiles, assuming that at 
large radii ($\sim$ 4"-7") the smooth flux is purely from the galaxy's stellar component. 
Since the isophotes should be the same shape at large radii, this 
last assumption provides a means of assessing the quality of our background subtraction. 
The background in the narrow-band images consists of 
three independent quantities: the bias offset, the sky flux, and the continuum. 
If the background is oversubtracted, the galaxy's isophotal profile
takes a characteristic sharp turn downwards at large radii. 
By minimizing this ``edge'', we were able to 
fine-tune our background subtraction for images with low signal to noise, 
and thus improve our ability to subtract the continuum from the narrow band images.

\begin{table}
\caption{\label{cal_dat}Molecular Line and Optical Line 
Parameters.}

\begin{tabular}{lllll} \tableline \tableline 

Target  & NGC1275 & A2597 &  & PKS0745-191 \\
Redshift	  			& 0.01756 	& 0.08520	&		& 0.1028 \\ 
\tableline
H2 Line Transition 			& 1-0 S(1) 	& 1-0 S(3) 	&		& 1-0 S(3) \\  
Observed Wavelength 	($\mu$)& 2.157		& 2.123		&		& 2.157 \\   
Continuum Filter	& F160W 	&  F160W	&		&  F160W \\ 
Line Filter			& F216N		& F212N		&		& F215N \\ 
Line Filter FWHM 	(\AA)	& 181.6	  	& 206.8		&		& 189.1 \\   
$\gamma_{c}$ (Equation 1)			& 1.09   	& 1.00		&		& 1.25 \\   
Nuclear Line FWHM 	(\AA)	& 50\tablenotemark{1} & 50\tablenotemark{1}	&		& 55\tablenotemark{2} \\   
Isophotal Counts Ratio  	  		& 0.056 	& 0.060 	&		& 0.055  \\   
Predicted Throughput Ratio		& 0.056 	& 0.056 	&		& 0.052  \\  
\tableline  

Optical Emission Line  		&H$\alpha$ &H$\alpha$ & OII           &H$\alpha$ \\   
Observed Wavelength 	(\AA)	& 6678		& 7122          & 4045	        & 7237 \\   
H$\alpha$/H$\alpha$+[NII]      & 0.44\tablenotemark{3}          & 0.41\tablenotemark{4}          &               & 0.38\tablenotemark{5}  \\   
Continuum Filter 		& F702W 	&  F160W	& F450W  &  F439W \\
Line Filter		& LRF			& F702W		& F410M		& LRF \\   
Line Filter FWHM 	(\AA)	& 77.7	  	& 1381.6    	& 219.4		& 84.1 \\   
$\gamma_{c}$ (Equation 1)			& 1.00   	& 1.08		& 1.09		& 1.00  \\   
Nuclear Line FWHM 	(\AA)	& 18\tablenotemark{3}	& 15\tablenotemark{4}	& 13\tablenotemark{4}	& 12\tablenotemark{5} \\   
Isophotal Counts Ratio 		& 0.041 	& 0.087		& 0.110 	& 0.387  \\   
Predicted Throughput Ratio 	& 0.043 	& 0.693\tablenotemark{*}	& 0.112 	& 0.169\tablenotemark{*} \\  
\tableline 
\end{tabular}
\tablenotetext{1}{Elston \& Maloney 1994.} 
\tablenotetext{2}{Falcke \etal  1998.}
\tablenotetext{3}{Unpublished Palomar Double Spectrograph spectrum (Donahue).}
\tablenotetext{4}{Voit \& Donahue 1995.}
\tablenotetext{5}{Donahue \& Stocke 1994.}
\tablenotetext{*}{The continuum and emission filters
were centered on significantly different wavelengths, thus the flat spectrum assumption breaks down.}

\end{table}

To test the reliability of this technique of scaling the continuum images, 
we used the synthetic photometry package SYNPHOT (available in STSDAS within
IRAF) to derive the relative flux 
scale an alternate way: by calculating the expected change in continuum throughput from filter 
to filter. Using the task {\em bandpar}, we were able to estimate the ratio of throughput efficiency 
for each filter pair for a flat spectrum  (in $F_\lambda$) source, 
where the efficiency is defined as the integral over wavelength
of the total filter throughput. These values are given in 
Table~\ref{cal_dat} (Predicted Throughput Ratio) and show
strong agreement with the isophotal predictions, except in the two cases (flagged with
asterisks) where the optical continuum 
filter is of a significantly different wavelength than the emission. 
In order to desensitize our final results to assumptions regarding 
the underlying continuum spectra, 
we based our continuum subtraction on the isophotal profiles. 

Because the effects of dust obscuration 
are much less significant at 1.6$\mu$m than at shorter wavelengths, 
we removed the continuum
from the WFPC2 emission-line images 
by fixing the isophotal parameters to match those parameters obtained from the 1.6-$\mu$m
data. For NGC1275, the bandpass of the 
companion broadband filter also contains the wavelength of the optical emission line 
of interest, but the filter is 
wide enough that contamination is only a few percent, and thus the 
net profile and line fluxes should not be significantly affected.

\begin{figure}
\plotone{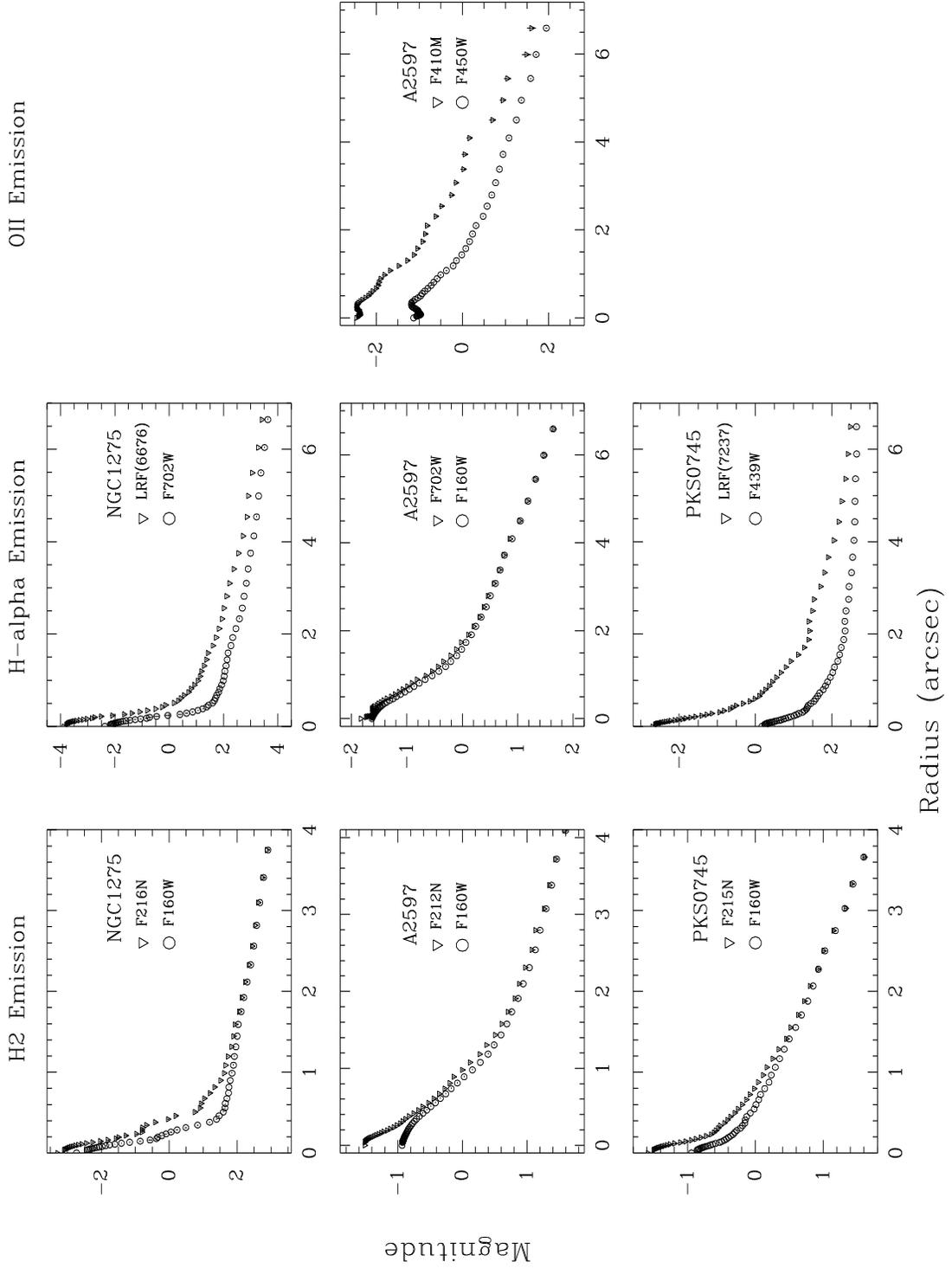}
\caption[]{Isophotal profiles of each source in the wide and narrow bandpasses.
\label{sb_profiles}}
\notetoeditor{Figure 1 would probably be best in landscape orientation.}
\end{figure}

We present the isophotal profiles for each galaxy in Figure~\ref{sb_profiles}, 
with magnitudes per unit sky area provided in 
arbitrary units. All of the NICMOS narrowband images  
clearly show H$_2$ line-emission within the 
central arcsecond of each cooling flow, and all three galaxies  
show additional extended 
emission, out to a radius of about 2-3". The optical continuum 
for all of the galaxies of course extends beyond the central 8".  
We present a morphological comparison of both the ionized and 
the vibrationally-excited molecular gas in \S\ref{maps}.

	\subsection{Absolute Flux Calibration \label{fluxcal}}
We describe here our process for estimating absolute fluxes for the 
emission line images. 
Once the residual emission images were created, we converted their count 
rates to units of absolute 
flux using the SYNPHOT package. 
For continuum observations of broadband sources, 
the standard HST flux calibration provides the 
correct absolute flux conversion with 
the overall photometric accuracy better than 5\% for the NICMOS data, 2\% for
the WFPC2 wide and medium band filters, and 3\% for the WFPC2 linear ramp filters.
A standard method to estimate the flux of a single 
emission line in ($\flux$) is to assume that the spectrum of the continuum emission is
flat (in $F_\lambda$), and then to correct for the contribution of the 
emission line. The following approximation is usually made for HST imaging: 
\begin{equation}
F_{line} = CR * \gamma_{c} * 1.054 * \Delta{\lambda} * PHOTFLAM 
\end{equation}
where $CR$ is the flux in DN sec$^{-1}$, $\Delta{\lambda}$ is the Gaussian 
width of the filter bandpass, PHOTFLAM is the flux conversion factor for
the standard HST flux calibration from
the HST data header (units of erg cm$^{-2}$ \AA$^{-1}$ DN$^{-1}$), 
and $\gamma_c$ is a correction factor to account for the position 
and width of the emission line in the filter bandpass ($\gamma_c=1$ 
when the line falls at the filter center)
(\anchor{http://www.stsci.edu/documents/data-handbook.html}{HST Data Handbook v3.1}, 
Section 18.2.5).  To estimate $\gamma_c$, we used SYNPHOT to model the
instrumental throughput for each filter, we assumed the emission lines 
to be Gaussian-shaped, and we used the H$_2$ line widths (FWHM) given 
in Table~\ref{cal_dat}. A DN is ``digital number'' or a count, and is
equal to the number of electrons read out from a pixel multiplied by 
the detector gain.

We tested the robustness of the standard method by
modelling a flat spectrum plus a single emission line  
first, then calculating the appropriate flux conversion.  
(See the \anchor{http://www.stsci.edu/instruments/wfpc2/Wfpc2_isr/wfpc2_isr9606.html}
{WFPC2 Instrument Science Report 96-06} describing these methods for the 
photometric calibration of the LRF's.) 
SYNPHOT convolves an emission-line
model of a given flux in erg sec$^{-1}$ cm$^{-2}$ and the appropriate Gaussian line width with the known throughput of the telescope 
in each relevant observing mode.  
SYNPHOT then returns the flux (in DN s$^{-1}$) 
that would be expected for each observation. 
Both methods discussed give similar results for a single Gaussian line,  
agreeing to within 5\% or less. 

However, each H$\alpha$ observation also included [NII] in 
its filter bandpass.  
The ratio of H$\alpha$ to 
the H$\alpha$+[N~II] emission for the central 2" of each cluster 
is reported in Table~\ref{cal_dat}, from ground-based spectroscopy.
To determine the net flux of H$\alpha$  emission, we had to 
estimate the contribution from
NII in each filter. For A2597, the F702W filter is sufficiently wide that the observed 
spectroscopic line ratios could be employed to correct the total measured flux 
for NII contamination. The linear ramp filters used to observe NGC1275 and PKS0745-191, 
on the other hand, are quite narrow and transmission falls off rapidly away 
from the filter center, and thus require a more sophisticated approach.  
To provide the most accurate estimate of pure H$\alpha$, we determined 
the flux calibration for the
specific case of the H$\alpha+$[N~II] line complex by 
convolving the appropriate filter and telescope response functions with 
three Gaussian lines centered at 6548\AA, 6563\AA, and 
6583\AA, having the appropriate relative flux ratios, linewidths, and redshifts 
measured from nuclear spectra
for each of the 3 galaxies (Table~\ref{cal_dat}). This 
correction increases the nominal flux conversion factor by ~10\% for NGC1275 
and PKS0745-191. We then corrected the total measured flux 
using the spectroscopic [N~II]/H$\alpha$  
line ratios (Table~\ref{cal_dat}).

The sensitivity of our flux calibrations to uncertainties in line widths and to 
the relative contributions of [N~II] and H$\alpha$ emission was also tested.  
The widths of the emission lines could be smaller for off-nuclear regions (H~II
regions, for example) than they are in the cores of the galaxies. 
The flux conversion factor for H$_2$ emission decreases by 8\%, 1\%, and 
20\% for NGC1275, A2597, 
and PKS0745-191, respectively,  if a FWHM of 5\AA~ rather than 50\AA~ is assumed.
For the H$\alpha$ line, decreasing the assumed FWHM by a factor of $\sim10$ to 
2\AA~ 
results in a 2\% flux decrease for NGC1275 and in no change for the other two targets.
The [N~II]/H$\alpha$ ratio assumed in our calculations is an average value 
from ground-based long-slit observations  
and could vary spatially. 
Varying the assumed [N~II]/H$\alpha$ ratio between 0.5 and 1.5 changes the 
total (H$\alpha$+[N~II]) measured flux very little, altering   
the total flux of NGC1275 and A2597 by 4\% and PKS0745-191 by 1\%. However,
the inferred H$\alpha$ flux could vary by a factor of 2.0 between these
two extremes, to be between 60\% and 33\% of the total measured flux. 
Therefore, off-nuclear pure H$\alpha$ estimates are only good to about a 
factor of 2. The absolute infrared H$_2$ flux calibration 
is good to at least 25\% off-nucleus where hot dust from the
central torus is unlikely to contaminate our measurements, and to approximately
a factor of 2 on-nucleus, lacking an estimate for the 
hot dust contribution. The 
uncertainties quoted in the following tables do not include these systematic
calibration uncertainties, but we take them into account when we 
compare these data to model predictions.

\section{Aperture Photometry \label{aperture}}

The derived isophotal profiles indicate excess two-micron emission in the 
central few arcseconds of each galaxy.  We know from previous spectroscopy
that each of the galactic nuclei produces H$_2$ emission, and that the 
continuum emission between 1.5 and 2.0 microns is fairly flat. So
our absolute photometry in the  H-band and 2-micron
narrowband filters should allow us to subtract the continuum in the nuclei
fairly accurately. We checked this by comparing the continuum scale factor 
derived from absolute photometry (\S\ref{isofit}) to that derived from 
matching the galaxy surface brightness profiles at the edges of the NICMOS
field of view. Nevertheless, as mentioned in \S\ref{fluxcal}, 
our estimates of H$_2$ emission in the nuclei, where
the AGN and hot dust from the central torus dominate, 
can not be precise because of the limitations on our technique.  
 
\begin{figure}
\plotone{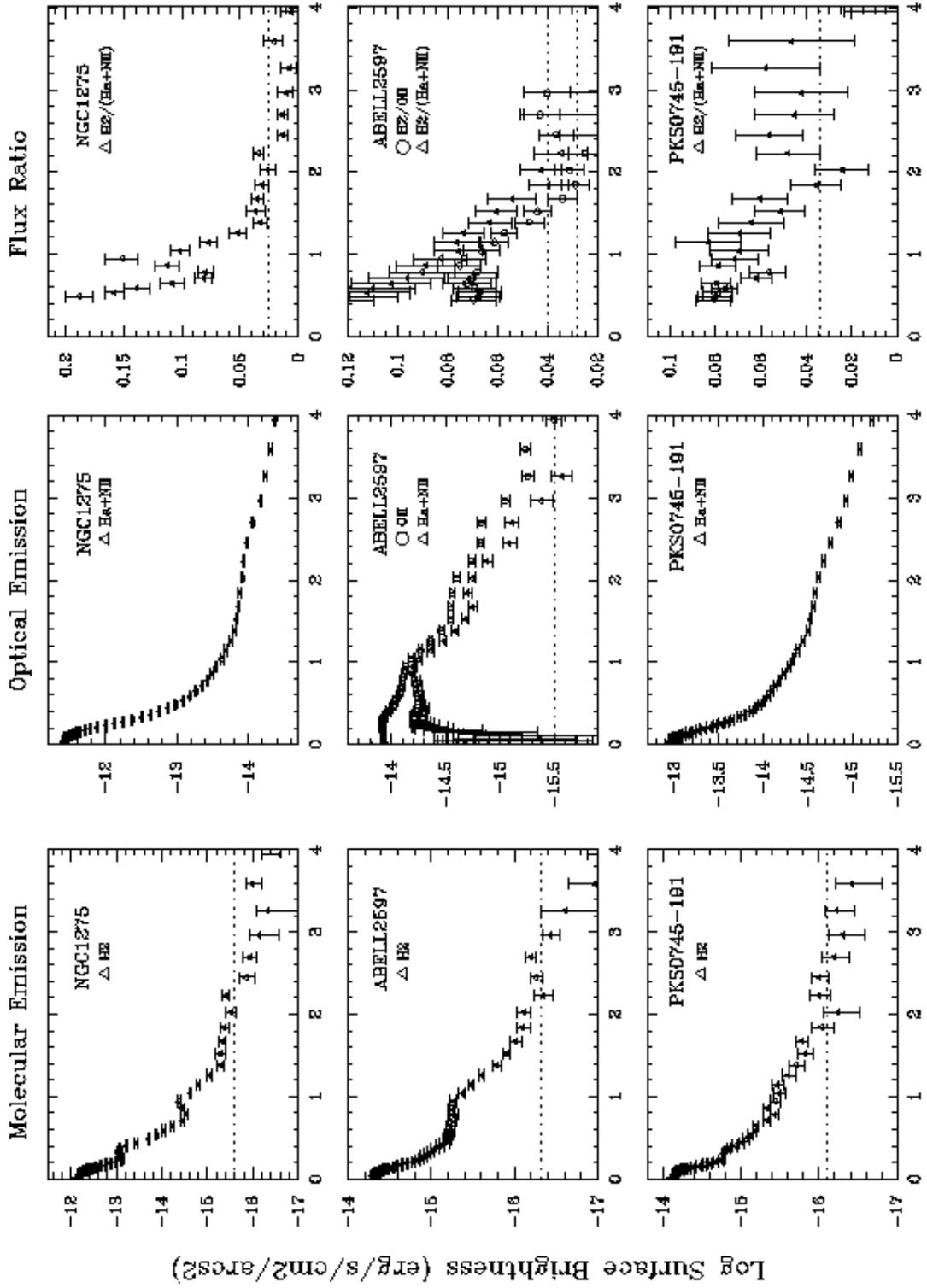}
\caption[]{Continuum-subtracted emission-line surface brightness and surface brightness 
ratios as a function of radius. The $1\sigma$ detection thresholds are plotted
with straight dashed lines. \label{line_profiles}}
\notetoeditor{Figure 2 would probably be best in landscape orientation.}
\end{figure}

The net H$_2$ surface brightness  
was measured from each residual image 
within concentric annular apertures centered on the galaxy. In 
Figure~\ref{line_profiles},  
the net surface brightness of the 
H$_2$ emission, the H$\alpha$+[N~II] (``optical'') emission, and the 
molecular to optical flux ratio for the inner 4 arcsec
of each galaxy is plotted. 
Spurious behavior of the flux ratios at radii less than 0.4" is an
effect of the NICMOS PSF and its Airy ring.   
The total H$_2$ fluxes, measured within a radius of 2" are 
presented in Table~\ref{fluxes}. 
The fluxes in these tables 
have been corrected for Galactic reddening and absorption but not for 
intrinsic reddening or absorption.

\begin{table*}
\caption{ Aperture ($r=2"$) Photometry Summary \label{fluxes}}

{\begin{tabular}{llll}\tableline\tableline\\
\multicolumn{1}{l}{} &
\multicolumn{1}{l}{NGC1275} &
\multicolumn{1}{l}{A2597} &
\multicolumn{1}{l}{PKS0745-191} \\
\tableline\\
Flux H$_2$       (DN s$^{-1}$)		& 176.6$\pm$6.02 &  7.09$\pm$1.23 &  6.73$\pm$1.21 \\ 
Flux $H\alpha+[NII]$(DN s$^{-1}$)	& 170.0$\pm$6.11 & 10.88$\pm$1.53 & 10.22$\pm$1.22 \\ 
Flux OII	 (DN s$^{-1}$)		&	         &  2.93$\pm$0.82 &		   \\
\\
Flux H$_2$          erg s$^{-1}$ cm$^{-2}$ & $(9.84\pm0.34)\times10^{-14}$ & $(3.71\pm0.64)\times 10^{-15}$ & $(4.63\pm0.83)\times10^{-15}$ \\ 
Flux $H\alpha+[NII]$ erg s$^{-1}$ cm$^{-2}$ & $(7.64\pm0.27)\times 10^{-13}$ & $(4.15\pm0.58)\times 10^{-14}$ & $(6.44\pm0.77)\times10^{-14}$ \\ 
Flux OII	    erg s$^{-1}$ cm$^{-2}$ &			         & $(5.87\pm1.66)\times10^{-14}$ &				\\ 
\\
Lum H$_2$  h$^{-2}$ erg s$^{-1}$ &  $3.3\pm0.1 \times 10^{40}$  & $3.0\pm0.5 \times 10^{40}$ & $5.5\pm1.0 \times 10^{40}$ \\
Lum H$\alpha$+[NII] h$^{-2}$ erg s$^{-1}$ & $2.6\pm0.1 \times 10^{41}$ & $3.4\pm0.5 \times 10^{41}$ & $7.7\pm0.7 \times 10^{41}$ \\
Lum [OII] h$^{-2}$ erg s$^{-1}$ &                               & $4.8\pm1.3 \times 10^{41}$ &  \\
\tableline
\end{tabular}}

\end{table*}

\textbf{NGC1275}:
For NGC1275, the molecular emission is concentrated in the nucleus 
of the galaxy. To determine its extent, we compared the radial profile of a 
typical star with the radial profile of the NGC1275 nucleus in the
narrow band image. The measured characteristic radius of the 
NGC1275 nucleus was  
only slightly larger 
than that of the PSF star, implying an intrinsic radial scale of less than 
0.09 arcsec ($\sim$1-1.2 pixels) for the residual emission, corresponding
to a physical scale of $\lta 22$ h$^{-1}$ pc. 
The H$_2$ 
emission line flux of the central source within a 2 arcsecond radius 
aperture was
measured to be $9.84\pm0.34 \times 10^{-14} \flux$, corresponding to an
H$_2$ line luminosity of $3.3 \times 10^{40} \lum h^{-2}$. Faint extended
emission is detected up to 2" off-nucleus.

The AGN in NGC1275 is known to be variable, so comparison of photometry
from different epochs may not be relevant, but we report earlier 
flux estimates here. Inoue et al. (1996) measured
the total flux of 2$\mu$m 
molecular emission in NGC1275 for a 2"x2" region in the center and derived a value
of $(2.5\pm0.1) \times 10^{-14} \flux$. Using a similar area 
aperture ($r=1.13"$), 
we measured $(9.4\pm0.3)\times10^{-14} \flux$.  
The ground-based Inoue et al. (1996) observations, done with a 2" by 30" slit, 
may not have been well-centered on the central source and the conditions 
were not perfectly photometric. 
Inoue et al. noted possible variable 
seeing and/or tracking problems with the standard star. Both problems would 
also affect flux estimates from a slit observation. They
also note a flux deficit with respect to earlier H$_2$ observations by 
Kawara \& Taniguchi (1993) ($3.8\times 10^{-14} \flux$ 
in the central 2.5" by 3.0"). Krabbe \etal  (2000) report ground-based
fluxes of $4.1\times10^{-14} \flux$ in the central 3". This discrepancy
suggests that our estimate could be contaminated by hot dust
continuum from the AGN, contributing almost equal flux  as the H$_2$ line 
to the  narrow-band excess on the nucleus. Our off-nuclear
surface brightness estimates are consistent with Krabbe \etal  (see
next section.)

In contrast to the compact vibrationally-excited molecular 
hydrogen emission, the H$\alpha$ line 
emission completely pervades the PC field of view. Optical 
line emission through the WFPC2 Linear Ramp Filter (LRF) 
is detected to the edge of the field; the precise surface brightness
of the extended emission is difficult to
quantify because the central wavelength of the LRF and the 
redshift of the emission-line gas varies across the field . The emission-line widths are relatively broad, and
the bandpass of the LRF at any position is relatively narrow,  
thus varying fractions of the [NII] and H$\alpha$ line complex are imaged,
depending on the position in the field. The location of the nucleus of 
NGC1275 is such that it is centered on the observed wavelength of H$\alpha$.
The H$\alpha$+[N~II] emission in an aperture of radius 2" is 
$7.6\pm0.3 \times 10^{-13} \flux$, and thus $F_{H\alpha}$ is  
$3.3\pm0.2 \times 10^{-13} \flux$. 
The ratio of H$\alpha$ to 1-0 S(1) H$_2$ line emission in the 
nucleus of NGC1275 is $3-7$ ($r<2"$, corrected for [NII] contribution and
for a contribution of warm dust continuum ranging from 0\% to 50\% of 
the estimated 2-micron excess in the nucleus. The ratio is  
uncorrected for internal absorption.)

\textbf{A2597}:
In A2597, the narrow-band infrared 
images reveal a complex, multi-component structure 
of vibrationally-excited molecular hydrogen extending 
over 2 arcseconds north and east of the nucleus. The location of the nucleus  
is revealed in the 1.6 micron image behind the thick dust absorption feature
seen in the optical images.
The complex structure of the emission gas is apparent in the differential flux 
profiles of A2597. The radial  surface brightness profile of [O~II]3727\AA~
decreases sharply at a radius of 0.3" where a thick dust lane obscures the galaxy's core. 
Within an aperture radius of 2", we derive a total H$_2$ flux of 
$(3.7\pm0.6)\times10^{-15} \flux$, corresponding to an emission line luminosity of $3.0 \times 10^{40} h^{-2}
\lum$.

We subtracted a scaled H-band image from the F702W image to reveal the 
emission-line system in H$\alpha$+[NII]. 
The H$\alpha$ image shows the same filamentary structure as does the
[OII] image (Koekemer \etal  1999), with an H$\alpha$+[N~II] flux of 
of $4.15\pm0.58 \times10^{-14} \flux$ within 2", or an
H$\alpha$ flux of 
$1.6\pm0.2 \times 10^{-14} \flux$  and an 
[O~II] flux of $5.9\pm1.7 \times10^{-14}
\flux$ inside 2", corresponding to optical emission 
line luminosities of $1.3 \times 10^{41} h^{-2} \lum$ 
and $4.8 \times 10^{41} h^{-2} \lum $ respectively. 
The ratio of H$\alpha$ to 1-0 S(3) H$_2$ emission in the 
central 2" is formally $4.3\pm0.9$, 
corrected for [NII] but uncorrected for internal absorption.

\textbf{PKS0745-191}: 
Emission from the H$_2$ gas within the central galaxy of PKS0745-191 
is concentrated in the core of the galaxy and in a
bright clump just off the nucleus. 
The total H$_2$ flux within a radius of 2"  (or 2.52 $h^{-1}$ kpc) 
is $(4.6\pm0.8) \times 10^{-15}$ erg s$^{-1}$ cm$^{-2}$, corresponding to
an 1-0 S(3) line luminosity of $5.5 \times 10^{40} h^{-2} \lum$. The 
total net H$\alpha$+[N~II] emission line flux from the same region is 
$4.6 \pm 0.8\times10^{-13} \flux$, or 
H$\alpha$ alone of $2.6\pm0.3 \times10^{-13} \flux$. 
The ratio of H$\alpha$ to 1-0 S(3) H$_2$ emission in 
the central 2" is $5.7\pm1.2$,
corrected for [NII] but uncorrected for internal absorption.

\section{Emission Line Maps \label{maps}}

\begin{figure}
\plotone{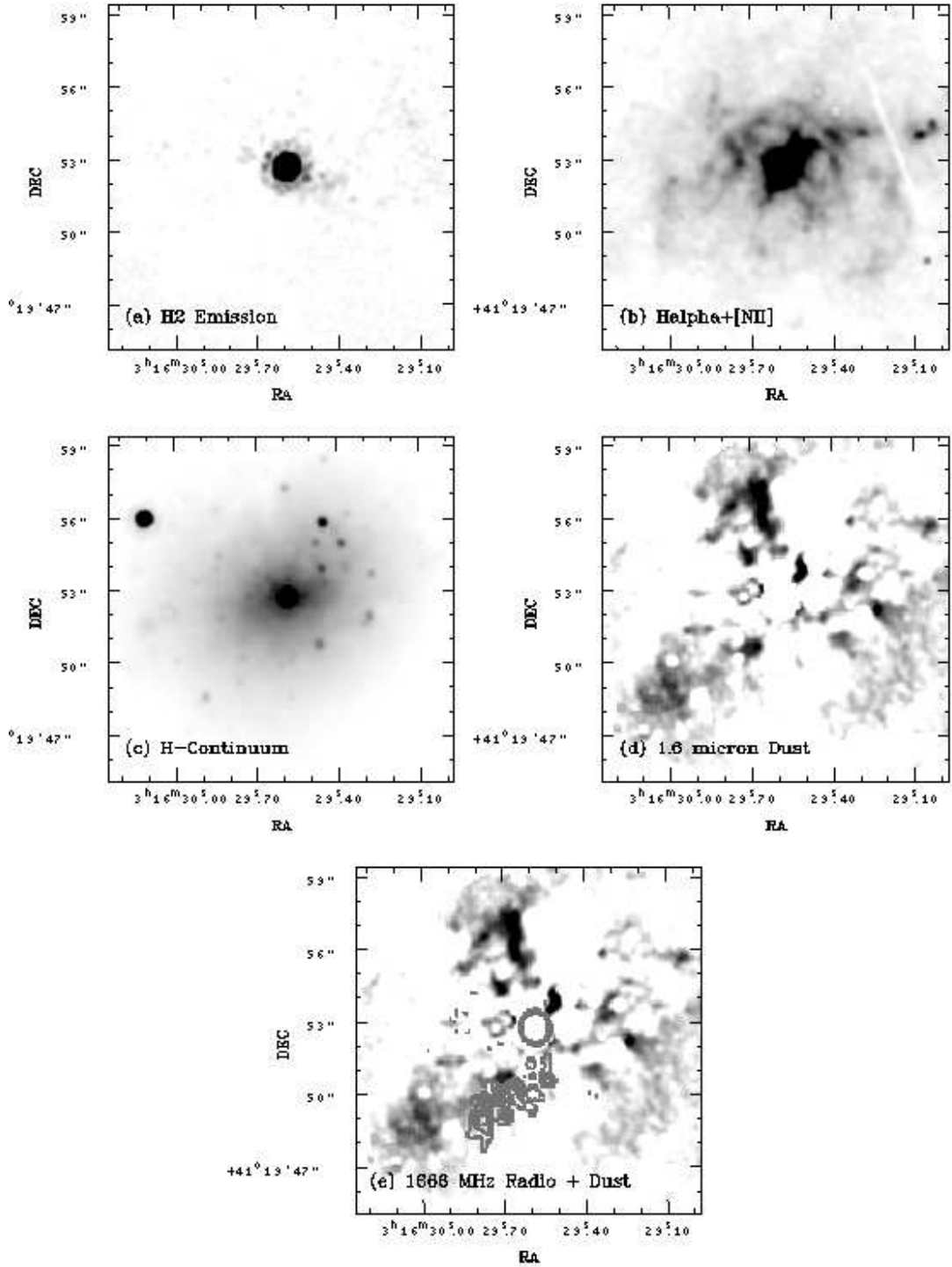}
\caption[]{Multifrequency images of NGC1275. (a) H$_2$ emission, (b) $H\alpha+NII$ 
emission, (c) H-band 
continuum, (d) 1.6 $\mu$m dust (darker regions are dustier), 
(e) 151 MHz Radio emission contours 
overlaid on a grey scale H$\alpha$+[N~II].\label{ngc1275}}
\end{figure}

\begin{figure}
\plotone{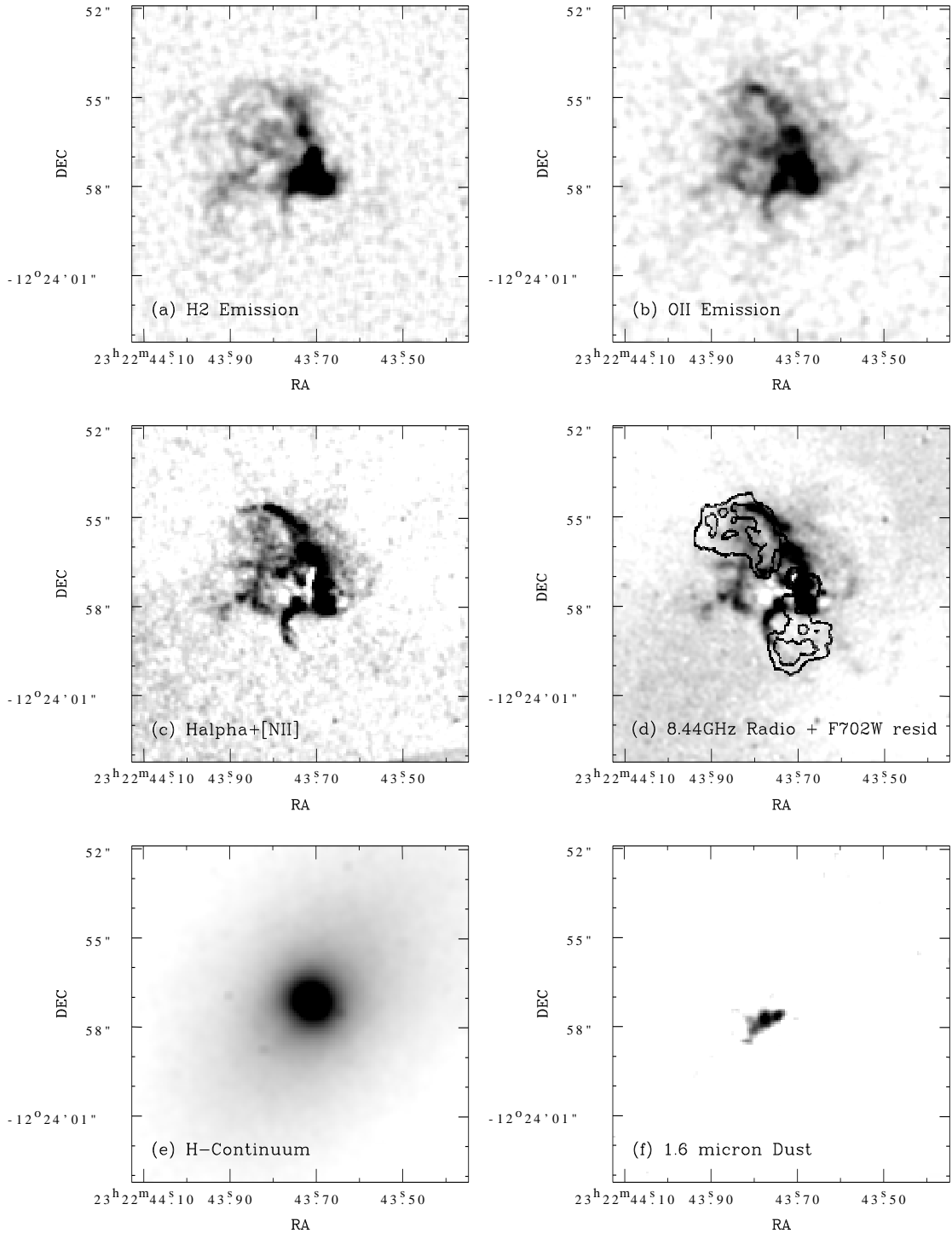}
\caption[]{Multifrequency images of A2597. (a) H$_2$ emission, (b) [OII] emission,
(c) H$\alpha$ +[N~II] emission, (d) 8.44 GHz radio emission (Sarazin \etal 1995) contours 
overlaid on a greyscale image of the F702W isophotal residual
image, emphasizing the blue lobes (seen here in white - negative) 
to the NE and SW of the radio source, 
(e) H-band continuum, 
(f) 1.6 $\mu$m dust (darker regions are dustier).\label{a2597}}
\end{figure}

\begin{figure}
\plotone{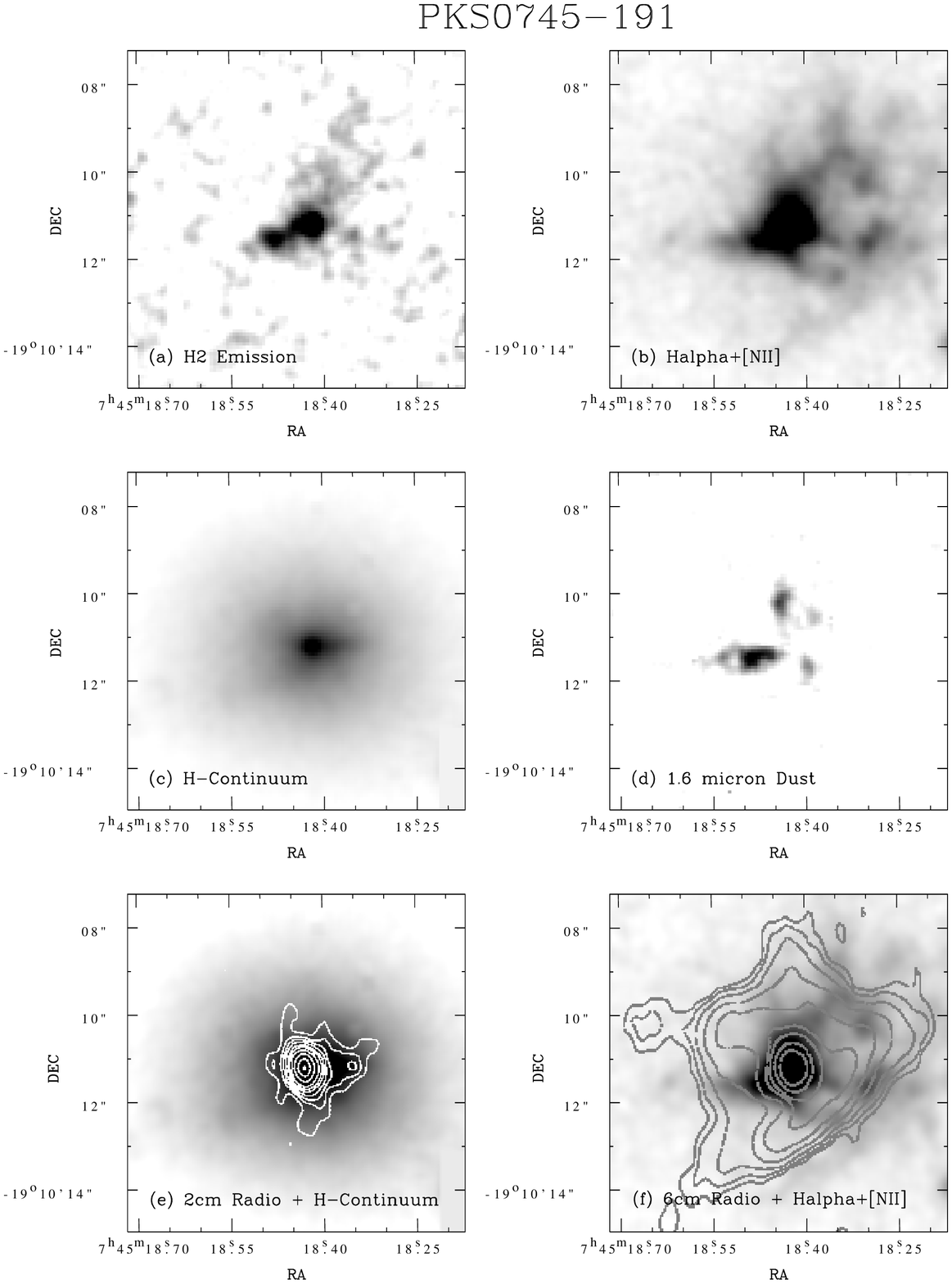}
\caption[]{Multifrequency images of PKS0745-191. (a) H$_2$ emission, (b) $H\alpha+NII$ 
emission, (c) H-band continuum, (d) 1.6 $\mu$m dust, (e) 2 cm radio contours from
Baum \& O'Dea (1991)  overlaid
on the 1.6 $\mu$m image, emphasizing the jet feature 
(f) 6 cm radio emission contours (Baum \& O'Dea 1991) overlaid on the H$\alpha$+[N~II] grey
scale image. \label{pks0745}}
\end{figure}

The morphology of the emission from the ionized gas, 
the vibrationally-excited molecular gas, and the dust absorption were compared for
each source. 
In Figures~\ref{ngc1275}-\ref{pks0745}, 
we present side-by-side grey scale comparisons of the molecular 
hydrogen emission and optical line emission along with the 
red starlight, the dust, and the radio emission for the central regions of each galaxy. We mapped 
the obscuration due to dust at 1.6$\mu$ by dividing each image by its isophotal model and binning 
the residual. For comparison with radio structures on the same angular scales, 
we sought radio maps from the literature 
with angular resolution similar to that of HST. Maps obtained at 
666 MHz for NGC1275 (Pedlar \etal  1990), 
at 8.44 GHz for A2597 (Sarazin et al. 1995), and at 2 cm and 
6 cm for PKS0745 (Baum \& O'Dea 1991) are contoured over HST greyscale images. 
All of the images of a given target 
are plotted at the same angular scale and orientation.

The emission line maps show what the aperture photometry and the radial
surface brightness plots only suggested: most of the 
molecular hydrogen emission from NGC1275 is confined to the nucleus, except
for faint wisps to the SW and NE, while the
emission from A2597 and PKS0745-191 is clearly extended. 
The maps reveal that this
extended emission is filamentary and traces very similar structures as does
the optical line emission. We report a possible jet feature in PKS0745-191. 
We now discuss each source in turn.

\textbf{NGC1275}: While extended, filamentary H$\alpha$ emission fills the 
central region of NGC1275 (Figure~\ref{ngc1275}b), 
we detected H$_2$ emission mostly 
in the nucleus of the galaxy (Figure~\ref{ngc1275}a). 
The H$_2$ residual image 
exhibits the pattern of the NICMOS PSF and its first Airy ring, 
with no structure detected at 
larger radii. There is a faint trail of emission off to the SW of the nucleus 
and a small blob east of the nucleus. These features were also seen in
ground-based images by Krabbe \etal  (2000). We measured similar surface
brightnesses. In a 1" by
1" box, the mean surface brightness of the trail feature is 
$1.5\pm0.8 \times 10^{-15} \sbr$, consistent with Krabbe's estimated
$2\times 10^{-15} \sbr$ for the same feature. 

The surface brightness limit for undetected features 
was computed by masking the obvious point sources 
in the residual H$_2$ image, and median filtering the residual within a smoothing kernel 
of an 13 by 13 pixel box. The mean net sky value was $0.00 \pm 0.005$ DN 
sec$^{-1}$ pixel$^{-1}$,
corresponding to a 2$\sigma$ surface brightness threshold of 
$5 \times 10^{-16} \sbr$.

H$\alpha$+[N~II] emission line surface brightness in the same region were as intense
as $2\times10^{-14}~ \sbr$ in a filament 3" west of the nucleus and $5 \times 10^{-14}~
\sbr$ in a bright feature extending 1" west-northwest of the nucleus. The total
amount of H$\alpha$ contributing to the emission-line flux is approximately 
half of the surface brightness in H$\alpha$+[N~II]. 
Therefore, in these bright optical 
filaments and features, the H$\alpha$/1-0 S(1) H$_2$ line ratio is greater 
than $\sim20$ and $\sim40$, respectively. These line-ratio limits are  
consistent with line ratios expected from shocked gas in which 
the molecules have
been largely dissociated ($V_{s} \gta 40$ km/sec).

The 1.6 $\mu$ image (Figure~\ref{ngc1275}c) shows a smooth stellar continuum and some dust
lanes, along with some of the stellar clusters
that were reported in WFPC2 optical observations by Holtzman \etal  (1992)
We report the IR photometry of the brightest
of these clusters in Appendix A. 
The exposure time for our GO image was only 
256 seconds, but this image was improved by co-adding an archival snapshot
image of 640 seconds.
The residual 1.6 $\mu$ continuum image (Figure~\ref{ngc1275}d) is displayed 
such that the regions with 
dust (negative residuals)
are grey or black, and the lighter areas are regions with 
less or no detectable absorption. 
The absorption map reveals that the central region is embedded in filamentary
dust features which wind inward as close as 0.5" from the center. The
features do not identically 
track the morphology
of the H$\alpha$ emission-line filaments. Some regions of strong dust absorption also
seem to have powerful H$\alpha$ emission, such as the region just NW of the nucleus, but
for the most part the dust features are at best tangled with the ionized gas filaments. 
The 666 MHz radio contours (Pedlar \etal  1990) show the AGN
point source and a feature extending SE parallel to a dust feature.

\textbf{A2597}:  The residual H$_2$ emission map (Figure~\ref{a2597}a) 
shows very distinct filamentary structure 
extending from the nucleus of the galaxy. This extended, filamentary network 
traces the complex structure found in both the H$\alpha$ and the [OII] emission maps 
(Figures~\ref{a2597}bc) derived from archival WFPC2 observations. Several bright 
filamentary arms extend outward from the nucleus and arc counterclockwise like tiny 
spiral arms or tidal tails. The 
brightest knot of H$_2$ emission is in the core of the galaxy. This has no direct 
counterpart in 
either of the optical emission images, likely because of substantial 
dust obscuration of the center of the galaxy. The optical emission 
profiles for Abell 2597 are
rather flat inside 1", and certainly not as peaked as the molecular line  
surface brightness, which may imply that dust may be absorbing the optical line emission at 
least in the central square arcsecond.

\begin{figure}
\plotone{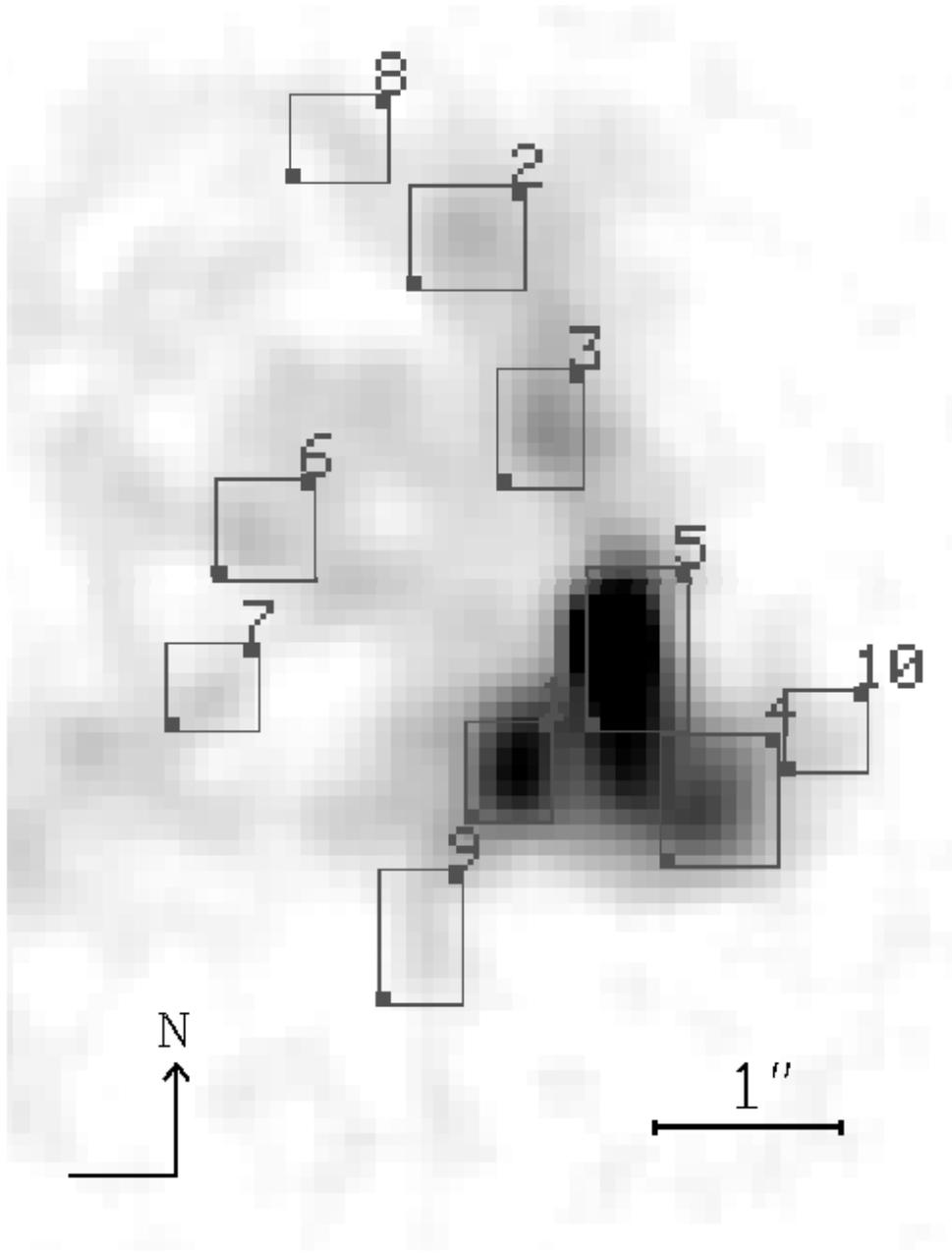}
\caption[]{A2597 Clump Regions\label{a2597_phot}}
\end{figure}

The 1.6 $\mu$m stellar continuum image (Figure~\ref{a2597}e), on the other hand, is 
quite smooth, with 
some very faint features which are hot pixels not
completely subtracted by processing. 
The residual 1.6 $\mu$m absorption image shows little dust 
absorption (Figure~\ref{a2597}f), except for one thick lane extending southeast from 
the nucleus. 
This dust lane is offset to the north of 
the nearby emission arm but is directly adjacent to it.
 The edges of the radio lobes traces the gas emission and the dust lane 
extremely well, with the leading edge of each radio lobe aligning precisely 
with a bright 
filamentary arm (Figure~\ref{a2597}d). 
These small-scale radio lobes are also correlated with the lobes of blue light, 
seen 
best in the F702W isophotal residual image (Koekemoer \etal  1999). 
The radio source appears to be interacting with 
the ambient gas and possibly also the dust. This interaction with the ICM may have induced star 
formation and would thus 
explain the presence of the blue lobes (McNamara et al. 1993; Koekemoer \etal  1999). 
In Figure~\ref{a2597_phot}, the brightest emission clumps
which were found in both the optical and infrared emission-line 
images are labelled, and we present the flux ratios of these
two components in Table~\ref{a2597_phot_table}. 

\begin{table*}
\caption{A2597 Individual Clump Fluxes \label{a2597_phot_table} }
{\begin{tabular}{lllll}\tableline\tableline
\multicolumn{1}{l}{Clump} &
\multicolumn{1}{l}{Area} &
\multicolumn{1}{l}{H$_2$} &
\multicolumn{1}{l}{$H\alpha+[NII]$} &
\multicolumn{1}{l}{[OII]}  \\
\multicolumn{1}{l}{} &
\multicolumn{1}{l}{arcsec$^2$} &
\multicolumn{1}{l}{$10^{-16} \flux$} &

\multicolumn{1}{l}{$10^{-15} \flux$} &

\multicolumn{1}{l}{$10^{-15} \flux$}  \\
\tableline
   1\tablenotemark{*}  &  0.236  & $2.91\pm0.09$ &   $1.69\pm0.02$ &  $2.42\pm0.06$\\
   2  &  0.315  & $1.13\pm0.03$ &    $1.70\pm0.01$ &  $1.69\pm0.04$\\
   3  &  0.270  & $1.47\pm0.05$ &    $2.10\pm0.01$ &  $1.81\pm0.05$\\
   4  &  0.405  & $3.5\pm0.1$ &      $4.71\pm0.03$ &  $4.62\pm0.13$\\
   5\tablenotemark{*}  &  0.433  & $7.0\pm0.4$ &     $3.79\pm0.02$ &  $5.29\pm0.12$\\
   6  &  0.276  & $0.85\pm0.02$ &    $1.03\pm0.06$ &  $1.31\pm0.04$\\
   7  &  0.203  & $0.49\pm0.02$ &    $0.829\pm0.004$& $0.93\pm0.02$\\
   8  &  0.236  & $0.43\pm0.02$ &    $1.44\pm0.02$ &  $1.41\pm0.05$\\
   9  &  0.304  & $0.63\pm0.03$ &    $1.28\pm0.01$ &  $1.28\pm0.05$\\
  10  &  0.203  & $0.52\pm0.03$ &    $0.426\pm0.006$& $1.04\pm0.03$\\

\tableline\\
\multicolumn{1}{l}{} &
\multicolumn{1}{l}{} &
\multicolumn{1}{l}{$H_2/[H\alpha+NII]$} &
\multicolumn{1}{l}{$H_2/[OII]$}  \\ \tableline
1\tablenotemark{*} & &      $0.172\pm0.017$    &   $0.120   \pm   0.005$\\
2 & &       $0.066    \pm    0.004$ &   $0.066   \pm  0.003$\\
3 & &       $0.070    \pm  0.003$ &   $0.081    \pm  0.003$\\
4 & &       $0.074    \pm    0.004$ &   $0.076   \pm   0.003$\\
5\tablenotemark{*} & &      $0.185   \pm  0.013$  &   $0.133   \pm   0.008$\\
6 & &       $0.083    \pm    0.004$  &   $0.065    \pm   0.003$\\
7 & &       $0.059    \pm    0.004$  &  $ 0.052    \pm  0.003$\\
8 & &       $0.030    \pm    0.003$  &   $0.031   \pm   0.002$\\
9 & &       $0.049    \pm    0.005$  &  $0.049    \pm   0.003$\\
10& &       $0.122    \pm    0.013$  &   $0.050    \pm  0.003$\\
\tableline\end{tabular}} 
\tablenotetext{*}{Regions with heavy 
dust obscuration which probably affects at least the observed 
optical line fluxes.}
\end{table*}

The core of the ROSAT HRI image of A2597 shows some east-west elongation at
the smallest scales resolvable by the HRI ($\sim4"$) (Sarazin \etal  1995; 
Pierre \& Starck 1998), which may indicate that the ICM of A2597 and the
ISM of PKS2322-12 (the central radio source A2597) 
are interacting within these scales.

\textbf{PKS0745-191}: Clumpy arms of H$\alpha$ emission spill from the center of 
PKS0745-19
(Figure~\ref{pks0745}b),
tracing the
emission from the H$_2$ gas (Figure~\ref{pks0745}a). The brightest molecular emission 
occurs in the core of 
the galaxy and 
in an adjacent knot to the southeast.  The H-band continuum 
image (Figure~\ref{pks0745}c) is very smooth and shows only a few patches of 
dust absorption in the negative 
residual image (Figure~\ref{pks0745}d). The most significant of the dust features
extends  
eastward from the core and lies directly opposite a linear continuum emission component which
appears to be a tiny jet. The 2 cm radio contours from Baum \& O'Dea (1991) 
(Figure~\ref{pks0745}e) shows a distortion in 
the same scale and position angle as the putative jet feature. 
Distorted and extended 6 cm radio contours, also from Baum \& O'Dea (1991),  
(Figure~\ref{pks0745}f) surround the core of the galaxy but exhibit no clear 
indications of 
jets or lobes. The shape of the 6 cm radio source suggests that some interaction may be
occuring between the system with the dust feature and the radio source.

\begin{figure}
\plotone{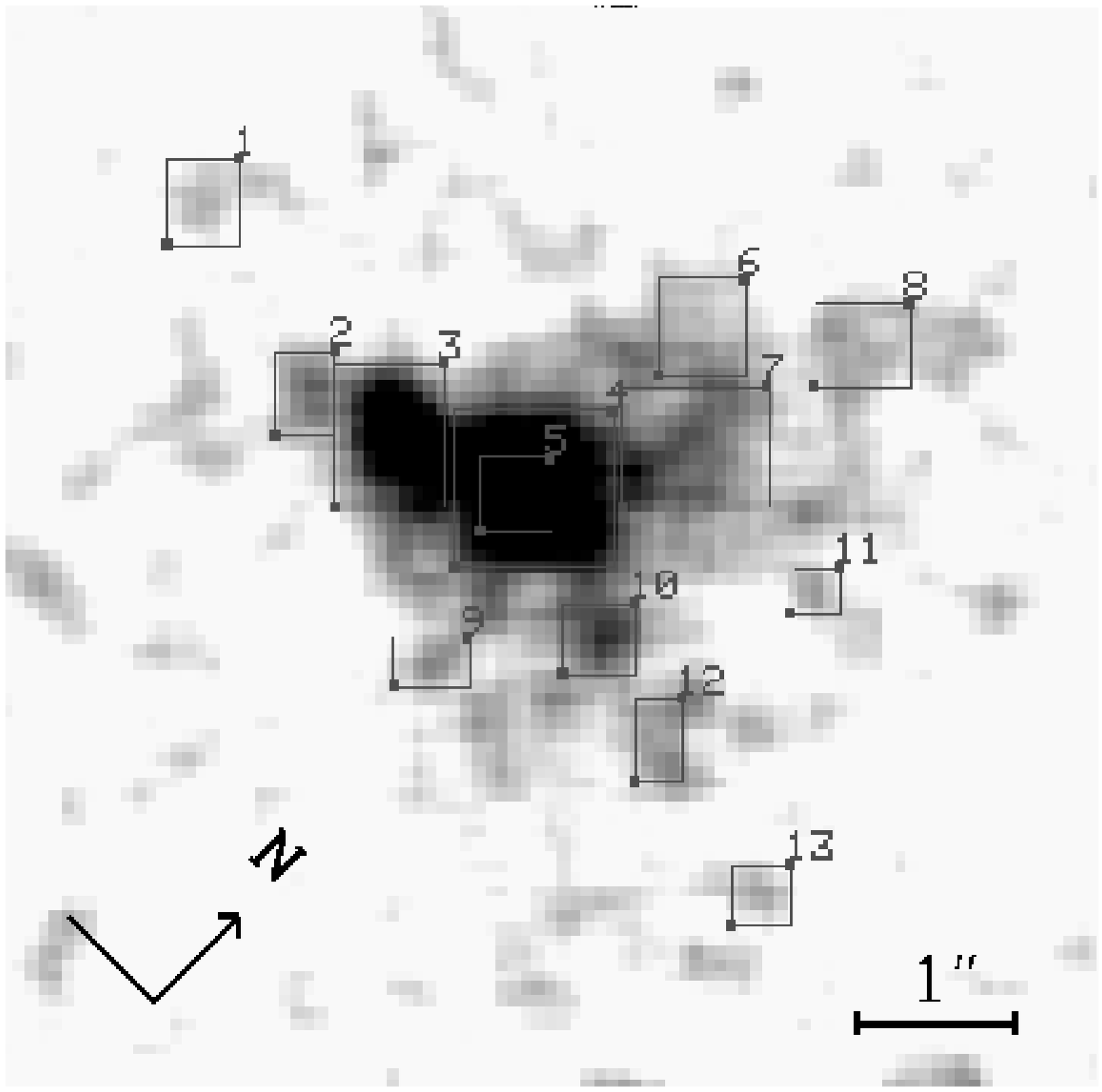}
\caption{PKS0745-191 Clump Regions\label{pks_phot}}
\end{figure}

In Figure~\ref{pks_phot}, we identify clumps of gas which are present in both the 
ionized and the molecular residuals, and we calculate the flux ratio of the two components for each 
clump in Table~\ref{pks_phot_table}.

\begin{table*}
\caption{PKS0745-191 Clump Fluxes. \label{pks_phot_table} }
{\begin{tabular}{lllll}\tableline\tableline
\multicolumn{1}{l}{Clump} &
\multicolumn{1}{l}{Area} &
\multicolumn{1}{l}{H$_2$} &

\multicolumn{1}{l}{$H\alpha+[NII]$} &

\multicolumn{1}{l}{$\frac{H_2}{[H\alpha+NII]}$}  \\
\multicolumn{1}{l}{} &
\multicolumn{1}{l}{arcsec$^2$} &
\multicolumn{1}{l}{$10^{-16} \flux$} &
\multicolumn{1}{l}{$10^{-16} \flux$} &
\multicolumn{1}{l}{} \\
\tableline
 1    &   0.315 &  $0.93\pm0.06$ &  $4.78\pm0.15$ &   $0.194\pm   0.015$\\
 2    &   0.270 &  $1.25\pm0.08$  &  $10.7\pm0.33$  &   $0.117\pm   0.008$\\
 3    &   0.731 &  $7.2\pm0.3$   &  $52.0\pm1.7$  &   $0.139\pm   0.008$\\
 4    &   1.100 &  $16.7\pm0.8$  &  $230\pm14$    &   $0.073\pm   0.006$\\
 5    &   0.276 &  $7.6\pm0.5$   &  $123\pm8$     &   $0.062\pm    0.005$\\
 6    &   0.405 &  $1.77\pm0.06$ &  $12.6\pm0.3$  &   $0.141\pm    0.006$\\
 7    &   0.804 &  $5.06\pm0.12$ &  $42.2\pm0.8$  &   $0.120\pm    0.003$\\
 8    &   0.405 &  $1.92\pm0.06$ &  $12.2\pm0.3$  &   $0.157\pm    0.006$\\
 9    &   0.197 &  $0.59\pm0.05$ &  $6.78\pm0.2$  &   $0.087\pm    0.008$\\
10    &   0.276 &  $1.60\pm0.07$ &  $13.1\pm0.3$  &   $0.122\pm    0.006$\\
11    &   0.141 &  $0.47\pm0.04$ &  $6.84\pm0.10$ &   $0.069\pm    0.006$\\
12    &   0.225 &  $1.08\pm0.04$ &  $12.0\pm0.25$  &  $0.09\pm    0.004$\\
13    &   0.203 &  $0.70\pm0.05$ &  $5.4\pm0.2$   &   $0.129\pm    0.010$\\

\tableline\end{tabular}}

\end{table*}

\section{Dust Extinction\label{dust}}

The optical and near-infrared images of all three galaxies show
significant dust lane features. These lanes show up as patchy or filamentary
absorption against a smooth elliptical distribution of light. 
By fitting and dividing out the elliptical backgrounds, the quantity of
absorption in each waveband can be estimated. The ratio of the absorption
in any two bands provides a point on the reddening law for the dust in 
that system. Using this method, we show that the dust in each of these
systems is consistent with a Galactic reddening law.

In the central region of NGC1275, dusty patches significantly obscure the 
galaxy, and this obscuration is still quite strong at infrared wavelengths
$\sim1.6\mu$m or H-band. A longer WFPC2 R-band  
exposure from the Hubble Data Archive provided the 
comparison opacity of the dust features at optical
wavelengths. (This exposure saturates the central
nucleus, so we did not use it to subtract continuum from the LRF
image.)  
If the underlying stellar emission is smoothly distributed across the
galaxy, the ratio of extinctions $A_R$ and $A_H$ 
could provide one point of the reddening law of the extinction. 
However, the emission maps created in the previous analysis revealed 
thick, clumpy patches of H$\alpha$  emission. Therefore optical line
emission contaminates the continuum estimate of this filter,  
and the extinction relation derived using 
this filter might be suspect. 
For this reason, we also use archival WFPC2 data at B-band (F450W) for the 
analysis of the dust lanes in 
NGC1275. Our conclusions using either the B or the R data with  
the H-band data are similar.

We forced the isophotal fitting parameters 
(the shapes and the centroids) of the optical images to match 
that derived for the NICMOS F160W image, except for the normalizations.  
We then divided the B-band, R-band, and H-band 
images by their isophotal models to create residual images which map the dust 
and allow a quanititative extinction estimate in magnitudes. The residual 
images were binned to $\sim0.2"$ resolution, and the detection limit for an
absorption feature was determined 
to be $\sim0.03$ magnitudes at H-band. Assuming the relations for Galactic-type 
opacity (Cardelli et. al 1989) and $R_v=3.1$, the H-band limit 
corresponds to a limit of 0.20 magnitudes at B, which we use 
for a lower cutoff. For reference, 
extinction ratios for each filter pair were calculated 
using the infrared and optical Galactic extinction relations with the mean 
filter wavelength (0.452, 0.687, and 1.593 $\mu$m at B, R, and H). This gives 
the result: $A_{F450W}/A_{F702W}=1.65$, $A_{F702W}/A_{F160W}=4.07$, and
$A_{F450W}/A_{F160W}=6.73$.

\begin{figure}
\plotone{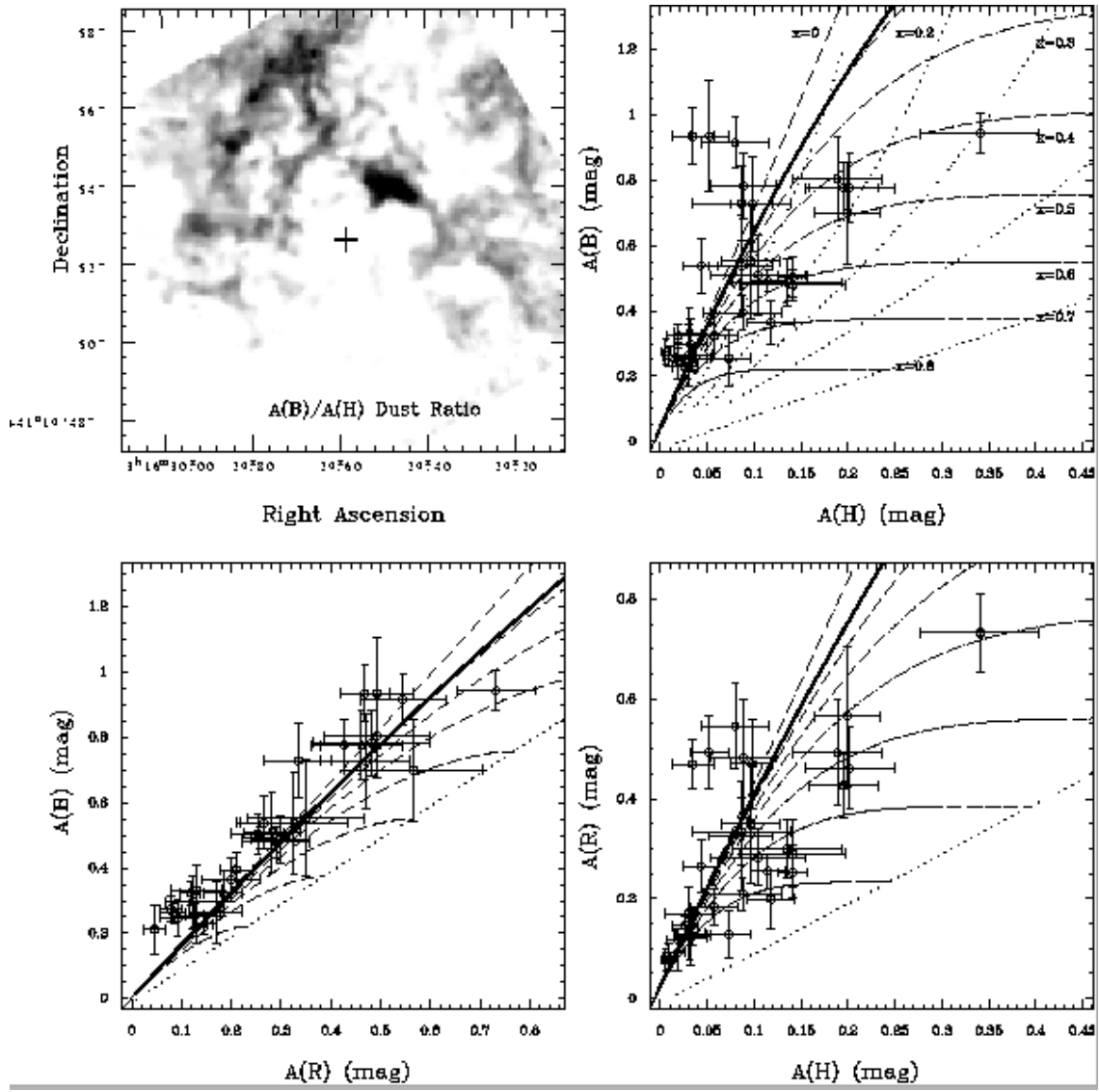}
\caption[]{Dust in NGC1275: (A. Upper left) Ratio of the extinction maps in the B-band
and the H-band. Black indicates a Galactic-normal reddening ($A_B/A_H=6.7$);
white indicates $A_B/A_H$=1.0 (either no or grey extinction). The cross marks
the center of the galaxy. (Note Fig.~\ref{ngc1275}d maps the absorption
features themselves.) (B. Upper right) $A_B$ vs. $A_H$
extinction, (C. Lower left) $A_B$ vs $A_R$. $A_R$ estimates are  
contaminated by filamentary [NII]+H$\alpha$  emission. 
(D. Lower right) $A_R$ vs $A_H$. 
The data points are the means and their rms deviations in 0.375" by 0.375" regions.
The bold line is the expected relation for a uniform mixture of
stars and dust. The dashed lines plot the relation when the dust is
geometrically thin, where $x$ is the fraction of light in front of
the dust. (See text for details.)
\label{ngc_extinction}}
\end{figure}

The image in Figure~\ref{ngc_extinction}(a) is a ratio map of the extinction 
magnitudes for the residual B- and H-band images, where black represents Galactic-type 
opacities ($A_B/A_H=6.7$), grey scales where the ratio is less than Galactic, 
and white indicates regions with no dust in either 
filter. We plot the measured extinction in two ways in Figure 
~\ref{ngc_extinction}(b): for each dusty pixel in the image and as a median extinction value, 
calculated in 0.1 mag bins. Two simple models described in Walterbos \& 
Kennicutt (1988) have been overplotted onto the data. The first of these (Model 
1, bold line) shows the expected relation for a uniform mixture of stars and 
dust. The second model (Model 2) assumes that the dust lane is geometrically 
thin and that the dust is embedded in a stellar disk, where $x$ is the fraction 
of light in front of the dust. Dashed lines represent varying values of $x$
(0.0-0.8), and dotted lines represent varying 
values of $\tau_h$ (0.2, 0.4, 0.6, 
1.0, $\infty$). For infinite optical depth, the expected extinction in the two 
colors is equal, since only light from sources 
in front of the dust would be detected.

In Figure~\ref{ngc_extinction}b, we see that Model 1 is an approximation to 
only some of the data. Model 2 
fits some of the points in a range of $0.4<x<0.8$ and $\tau_h<0.4$. 
In the image, these points correspond mainly to the large dust feature  
northeast of the galaxy's center, where the 1.6 $\mu$m opacity is the highest (see Figure 
5d).  The correspondence to Model 2 implies that the dust in 
the central few arcseconds of NGC1275 is well-mixed with the stars, but 
not in a uniform manner. Optically thick patches appear to be intermixed with 
regions showing Galactic-type opacities.
These regions are the darkest regions in the grey scale 
Figure~\ref{ngc_extinction}(a).

In Figure~\ref{ngc_extinction}c and ~\ref{ngc_extinction}d, we plot 
the $A_B$ vs. $A_R$ and $A_R$ vs. $A_H$ extinctions. 
Contamination due to optical line emission at R-band or B-band changes
the intercept of the color plots.  
Certainly some contamination occurs in the broad R-band filter. Assuming 
that contamination is $7\%$, this effect moves data points in 
Figure~\ref{ngc_extinction}c to the 
right and in Figure~\ref{ngc_extinction}d up  
by 0.07 magnitudes. A straight-line fit to the $A_B$ vs $A_R$ 
extinction, allowing for error bars in both directions, 
reveals a slope of $1.45\pm0.13$, somewhat less steep than a simple
screen of  
Galactic dust, but consistent with some mixing between stars and dust. 
A formal intercept of $0.10\pm0.03$ magnitudes is found, which
is consistent with our estimate of the contamination in R.
Three outlying points, with apparently high extinction in the blue
band but very little extinction in the R and H band, lie very near a 
strong dust feature just NE of the center of the image. Inside this
feature lies a very red star cluster (perhaps a globular cluster) which
is strongly absorbed at B, but is just visible in H.

The most significant result to be noted from the absorption
comparison plots are that the slope of the extinction relation appears 
to be either the same as, or less steep than, that of a Galactic extinction
relation, which is consistent with a heterogeneous mix of stars and 
Galactic dust. Therefore, we see very little evidence for an unusual 
dust law in NGC1275. There is a significant amount of dust, which 
we can see in distinct features even at
1.6 $\mu$m, but the reddening is consistent with that of 
normal Galactic-style dust. The NGC1275/Perseus observations 
do not show the unusually steep 
slope for the absorption ratios as measured for Abell 1795 by Pinkney et al (1996). 

In Abell 2597,  spectroscopic determination of
$A_R$ from the Balmer line decrement and the assumption of a standard
reddening law in the central two arcseconds 
provided an estimate of 
an absorption of $A_R\sim1$ (Voit \& Donahue 1997).
Estimating the dust absorption and reddening from existing HST imaging
for Abell 2597 is problematic. 
We have no line-free optical continuum image for Abell 2597. For the 
strongest absorption feature in the image, a triangular aperture wedged 
alongside the nucleus, $A_R \sim 0.10\pm0.01$ magnitudes 
and $A_H = 0.009\pm0.002$ magnitudes, giving a ratio of
$10.7\pm2.8$.  This ratio, however, must be suspect. 
The red filter measurement is strongly contaminated by 
line emission from the H$\alpha$+[N~II] complex (so much so we can use the
image to estimate H$\alpha$ fluxes).  The red ``absorption'' of this feature 
may be wrong because our technique for estimating it is not
robust to significant amounts of structure, which certainly exists nearby
the nucleus of A2597. The observed ratio in this small region is nonetheless 
consistent with what is expected from optically thin 
Galactic-style dust, but high resolution, high quality 
pure optical continuum observations or deep optical spectroscopy are 
needed for a reliable measurement of dust. 

The continuum (F439W) image of PKS0745-191 also does not permit a 
very accurate
measure of the color sensitivity of the absorption in the single dust 
feature in that source.  In the dust 
bar feature just west of the nucleus, $A_B =0.16\pm0.04$ 
and $A_H= 0.05\pm0.01$ magnitudes, corresponding 
to a ratio $2.9 \pm1.1$, less than the expected Galactic ratio of 6.7, 
but consistent with a dust slab where 80\% of the observed starlight lies in
front of it.

In summary, we have unambiguously 
detected the absorption of dust features in all three galaxies. 
For NGC1275, we have extensive measurements for the 
prominent dust features in that system, 
while for A2597 and PKS0745-191, the estimates were possible for 
only for a single feature in each system. 
For all three sources, the ratios of the optical to the
infared absorption are consistent with Galactic
reddening. The dust absorption is consistent
with that of rather thick, normal dust clouds lying behind 
foreground stars.

\section{Origins of the Gas \label{source}}

In order to emit in the 1-0 S(1) or 1-0 S(3) vibrational 
transition line, molecular hydrogen must be vibrationally excited. If it
is excited collisionally, it must be relatively warm ($1000-2000$K). 
To infer anything more about this gas, we
would like to know where it came from and how it is heated. In the 
following section we discuss the possible origins of the gas and
of the gas morphologies.

\subsection{The molecular gas is not part of a cooling flow}

The molecular emission lines are far too bright to be from cooling gas
(Jaffe \etal  1997). If molecular line emission is responsible for  
radiating energy from a cooling flow as the gas cools through the temperature 
2000K, the fraction of cooling expected from a single H$_2$ infrared 
lines is about $\eta = 2-10\%$. Estimating the mass cooling rate by 
\begin{equation}
\dot{M} = \frac{m_{H_2} \Phi}{\eta}
\end{equation}
where $m_{H_2}$ is the mass of a hydrogen molecule and $\Phi$ is the
line luminosity in photons per second, we find a mass cooling rate inside
the central 2" of tens of thousands of solar
masses per year, two orders of magnitude larger than the {\em total} cooling
flow rates over an entire cluster (Table~\ref{targets}). 
Specifically, the mass cooling rates
implied for NGC1275, A2597, and PK0745-191 would be 18,000, 16,000, and
29,000 $h^{-2} M_\odot$ yr$^{-1}$ respectively.

One possible source of the gas, however, could be the intracluster medium 
(ICM). It is difficult to imagine scenarios where molecule formation would
occur readily in the cores
of cluster cooling flows (Voit \& Donahue 1995), although the possibility
has been explored (Ferland, Fabian \& Johnstone 1994). 
Molecular hydrogen formation occurs most 
readily when dust is present (although there are formation scenarios where
dust is not required). Since dust shouldn't last long in the hot ICM, we
don't expect molecular hydrogen to form in gas that has condensed out of
the ICM. Furthermore, the large energy density in X-rays will inhibit both dust
and molecule formation (Voit \& Donahue 1995). 
Therefore, this molecular hydrogen is unlikely to be the end 
product nor direct evidence of a cooling flow.

\subsection{Morphologies of the dust and optical and infrared emission-line gas}

The emission-line morphologies in these clusters indicate that the
H$_2$ gas is closely related to the H$\alpha$ emitting gas. 
The emission-line filaments of H$_2$ align nearly exactly with the 
emission-line filaments of H$\alpha$. We are unable to 
tell whether the correspondence
extends to lower surface brightnesses, but the highest surface brightness features
seem to overlap substantially.  The correspondence of the optical and 
infrared emission-line filament structures suggests, but does not demand, 
that the heating mechanisms might also be related. 

The filaments themselves are concentrated
in the central region in bright fuzzy blobs and extend away from the nucleus
in delicate tendrils that seem to surround cavities that may have been evacuated
by radio sources. Some of the filaments have knots which could
be simply brighter parts of the filaments or HII regions which are photoionized
by stars. The structure of the extended filament system is suggestive of
shocks; however, we know from nuclear emission-line 
spectroscopy of A2597
(Voit \& Donahue 1997) and from stellar continuum analysis of central
galaxies in cluster cooling flows 
(Cardiel, Gorgas, \& Aragon-Salamanca 1998b) that
hot stars play a significant role in the optical line emission from
these galaxies.

The arcsecond-scale radio sources in both Abell 2597 and PKS0745-191, and,
to a limited extent, in NGC1275,  appear
to be interacting with the  atomic and vibrationally-excited molecular gas.  
The radio plasma in these galaxies seems to be affecting and 
is affected by the atomic and molecular gas.

The dust absorption features of these three 
galaxies do not align with the emission-line 
filaments although there is some overlap. The dust features tend
to be elongated in the same direction as the emission-line feature, and
in some cases, the dust lanes run alongside the emission-line filaments. 
The conclusion of this morphological observation is 
that while the
emission-line gas itself is known to be dusty based on significant
calcium depletion -- that is, no or very little [CaII] 
emission has been observed (Donahue 
\& Voit 1993), there is certainly 
dusty gas which is not associated with the emission-line filaments, and
may be quite distinct from them.  
This dust seems to be quite normal in NGC1275, A2597, and PKS0745-191 
as we showed in \S\ref{dust}. Sparks, Ford \& Kinney (1993) 
and Sparks, Macchetto, \& Golombek (1989) discovered
normal Galactic-style dust in M87 and NGC 4696 (Centaurus) respectively.
(In these systems, the emission-line gas and the dust appear to be
co-spatial, at least on arcsecond scales.)  
If the dust properties are consistent with those of Galactic dust, 
it is unlikely that the dust has   
been processed by X-rays or is a byproduct of any  
process whereby the nebular or dusty gas condenses from the hot gas.

It is therefore unlikely that the dust, the ionized emission-line gas, and by association, the warm
molecular gas, originated as intracluster material at X-ray emitting
temperatures. Sparks et al (1989, 1993) have speculated that a merger
event provided the gas, citing the coherent filamentary structure, the
quantity of dust, and the angular momentum of the emission-line gas 
as evidence.

\section{Heating the Gas \label{heat}}

Our observations allow us to test hypotheses about how the molecular
gas might be heated. 
Vibrationally-excited molecular hydrogen is present in 
both active galaxies and in starburst
galaxies. Therefore, both stars and AGN can vibrationally
excite molecular hydrogen. We note that 
the exact physical process responsible for the excitation is not
clear even for starbursts and active galaxies.

The molecular emission from NGC1275 is likely to be dominated by its AGN
but that of Abell 2597 and PKS0745-191 is not. Therefore our
discussion in this paper will focus on testing explanations for what
is heating the molecular gas in these two systems. 
The point-like morphology of the H$_2$ emission of NGC1275 suggests 
that it is probably
related to nuclear activity. 
Other Seyferts show nuclear 
infrared H$_2$ emission on small size scales. 
NICMOS imaging of nearby Seyfert galaxies by Quillen \etal  (1999) 
revealed resolved or extended 2-$\mu$m hydrogen line
emission from 6 out of 10 Seyferts, coincident with the H$\alpha$+[NII] 
line emission, and on much smaller physical scales than we observe 
in Abell 2597 and PKS0745-191 (several
100 pc rather than a few kpc, as observed in A2597 and PKS0745-191). 
The  H$_2$ to H$\alpha$ line ratios observed by Quillen \etal (1999) 
are similar to that of NGC1275.

In contrast, the H$_2$ to H$\alpha$ ratios in both the nuclei and in the
off-nuclear filaments in A2597 and PKS0745-191 are unusually 
high, even with the significant uncertainties that are incurred in 
correcting for [NII] emission and other systematics. 
Extinction of H$\alpha$ relative to H$_2$ may be occurring, and this 
certainly appears to
be the case just eastward of the A2597 nucleus where there is a significant
dust feature, H$_2$ emission, and no H$\alpha$. But global source-wide extinction 
does not appear to be a significant factor in producing the high ratios.
(Observations of A2597 Balmer lines suggest reddening consistent with 
$A_v\sim1$ (Voit \& Donahue 1997).)

Falcke \etal  (1998) finds  
an H$_2$ to P$\alpha$ ratio of 0.50 for PKS0745-191 and 2.02 for 
A2597. If we scale our H$_2$ observations by these
ratios, we can compare this rough estimate of the 
H$\alpha$/P$\alpha$ ratios to that expected for  
Case B H$\alpha$/P$\alpha$. For $T_e=10^4$K, Osterbrock (1989) 
predicts 8.3 (see also Brocklehurst 1971), and the ratio 
ranges between 7.5 - 9.7 for $T_e=5,000-20,000$K. 
Our observed nuclear H$\alpha$/P$\alpha$ ratios
are $8.7 \pm 1.8$ for A2597 and $2.9 \pm 0.6$ for PKS 0745-191.
The ratio in A2597 is consistent with Case B recombination with
little or no extinction. We note that a far more reliable analysis of the
Balmer series for a single deep longslit 
observation of A2597 by Voit \& Donahue (1997) 
obtains $A_V \sim 1$, without the aperture uncertainties inherent
in this comparison. The  H$\alpha$/P$\alpha$ ratio
in PKS0745-191 is consistent with a differential absorption  
between H$\alpha$ and P$\alpha$ of only 50-60\% at the wavelength of  
H$\alpha$, corresponding to a modest $A_V\sim1$ (Hill, Goodrich \&
Depoy 1996; Seaton, 1979). We conclude that the internal extinction of the
emission line gas is not huge, $A_V\sim1$ for both sources.

We will thus use the following three pieces of evidence to test the physical
 models, supplementing with observations at UV, radio, and X-ray wavelengths
as appropriate:
\begin{enumerate}
	\item The morphological similarity between the emission line gas
		in the optical and in the near infrared for A2597 and PKS0745.
	\item The unusually high intrinsic 
		ratios of H$_2$ emission to H$\alpha$ 
		emission, both globally and locally in A2597 and PKS0745.
	\item The relatively intense surface brightness of H$_2$ emission
		as measured locally in A2597 and PKS0745.
\end{enumerate}

\subsection{X-ray Photoelectric Heating, Conduction, and Mixing Layers}

X-rays can be quickly ruled out as a viable universal 
source of heat for the near-infrared filaments in at least one
of the clusters, Abell 2597. X-ray photoelectric 
heating can excite H$_2$ vibrational lines deep within the cloud. It is 
a process which can reproduce the H$_2$ to H$\alpha$ line 
ratios, given a sufficient column depth of hydrogen. X-ray irradiation and
heating will also destroy H$_2$, but warm ($T\sim 1000-2000$ K) H$_2$
radiates so efficiently in the 1-0 S(1) line 
that only a small column of H$_2$ ($\sim3\times10^{18} 
\, {\rm cm}^{-2}$ out of a total hydrogen column of $\sim10^{21} 
\, {\rm cm}^{-2}$, a molecular fraction of only $\sim3\times10^{-3}$) is 
required to produce the observed surface brightnesses.

We used archival ROSAT HRI images to assess the local X-ray flux in 
each source. The ROSAT HRI detector acquires a simple image  (no spectra) 
of an X-ray source between $0.2-2.0$ keV, with a spatial resolution
of $\sim4"$.  
The $0.5-2.0$ keV X-ray luminosity as measured from
HRI imaging within a circular aperture of $r= 8$" and a $4 \times 4$" box 
for each target is listed in Table~\ref{xray}. We corrected the observed
X-ray luminosities to 
bolometric X-ray luminosities by assuming the temperatures and Galactic columns
(Table~\ref{targets}) and 30\% solar metallicities. 
The ratio of 
the bolometric X-ray flux to the H$_2$ line flux shows a wide variation, from
7 for A2597 to 400 for PKS0745. We did not attempt to correct the 
bolometric X-ray flux to H$_2$ line flux ratios for the different sizes of
apertures in the two measurements. (Higher spatial resolution images from 
Chandra would provide a
better direct comparison, of course.)

\begin{table}
\caption{X-ray Measurements from ROSAT HRI \label{xray}}
\begin{tabular}{lcccc} \tableline \tableline
Name    & $L_{0.5-2.0~{\rm keV}}$ ($r<8"$) & $L_{0.5-2.0~{\rm keV}}$ (4" $\times$ 4" box)  & Bol. Corr  & $L_{bol}/L_{H_2}$\\ \tableline
NGC1275 &$1 \times 10^{42} h^{-2} \lum $& $1.8 \times 10^{41}h^{-2} \lum$ & 4.22
	& 23 \\
A2597   &$5 \times 10^{41} h^{-2} \lum $& $5.5 \times 10^{40}h^{-2} \lum$ & 4.12
	& 8 \\
PKS0745 &$2 \times 10^{43} h^{-2} \lum $& $2.5 \times 10^{42}h^{-2} \lum$ & 8.01
	& 360 \\ \tableline
\end{tabular}
\end{table}

For gas that has been heated to $\sim2000$ K, the 1-0~S(1) line
emits about 0.5\% of the absorbed X-ray luminosity. This line is responsible
for about 10\% of the total H$_2$ line flux, which is in turn responsible
for about 10\% of the total cooling in an X-ray heated cloud (e.g.
Lepp \& McCray 1983). About 50\% of the absorbed X-ray
energy goes into heating the gas (in contrast to photodissociation 
regions where the efficiency
is closer to 0.1\%). Typical X-ray/S(1) ratios obtained from X-ray irradiation
 models (Maloney, Hollenbach \& Tielens 1996) 
are around 2000, since only about 10\% of the incident flux
is absorbed by clouds with $\sim10^{22}$ cm$^{-2}$ columns. The 
ratio can be smaller for a very steep incident spectrum. 

In the case of NGC1275, the AGN may be supplying a significant fraction
of the central X-ray luminosity, and therefore, with a steeper spectral 
component, the predicted ratio between X-ray and 1-0~S(1) flux is lower and
closer to that observed. But for Abell 2597,  
significant X-ray heating of the H$_2$ gas seems highly unlikely given
the absence of a significant intrinsic
absorption column. Its $L_x/L_{H_2}$ ratio of 7 within the $4" \times 4"$
aperture is far too low for
X-ray heating to be energetically possible.  
PKS0745-191 is closer to the Galactic plane, and 
therefore its Galactic column and the uncertainty on its intrinsic 
column is higher. If PKS0745-191 has significant intrinsic absorption,
X-ray heating of its H$_2$ gas is remotely possible but not likely. High signal-to-noise 
Chandra CCD observations would have sufficient angular resolution to test
whether there is significant intrinsic absorption of X-rays, using spectra 
from only the central few arcseconds.

Based on relative observed fluxes, while 
the local X-ray flux compared to the optical and molecular emission line
flux is insufficient to produce the line emission by X-ray heating for
Abell 2597, photoelectric heating of the molecular gas 
by X-rays is remotely feasible in the cluster PKS0745-191.
To explore what H$\alpha$/H$_2$ S(1) line ratios one might expect from 
X-ray heated gas, 
we have constructed 
a constant pressure model, using the methods described
in Maloney, Hollenbach, \& Tielens (1996), where $P/k = 10^8$ cm$^{-3}$ K, 
the  incident spectrum is flat (in $f_\nu$) with
a cutoff at 8 keV to mimic a thermal spectrum, and a total cloud
column of $1.5 \times 10^{22}$ cm$^{-2}$. 
A warm ($T \sim 8000$K), 
weakly ionized ($x_e\sim 0.01-0.1$) atomic zone is produced near
the surface; with increasing column density into the slab the gas
undergoes a transition to a molecular phase.
The ratio of H$\alpha$ to S(1) (including
H$\alpha$ from the warm atomic zone only) is about 4; including the
H$\alpha$ contribution from the highly ionized layer at the surface of
the slab raises the ratio to 8, comparable to the ratios we observe in
these clusters.

We note that the H$\alpha$/S(1) ratio is sensitive to the slope
of the continuum below a few hundred eV and the total cloud column density. 
If the spectrum is much softer than what we assume (that is, if 
the number of photons between 13.6 and 100 eV is significantly
larger than the number of photons between 1 and 8 keV), the H$\alpha$ 
contribution from the fully ionized layer could be much larger. 
Also, if the total column density of the cloud is $5 \times 10^{21}$
cm$^{-2}$ rather than $1.5 \times 10^{22}$ cm$^{-2}$, the molecular
gas column decreases and the predicted H$\alpha$ to S(1) ratio increases
to 80.

Therefore, for X-ray heated gas, the predicted 
H$\alpha$ to S(1) line ratios can be tuned to a range of values 
consistent with what we observe. If these clouds are X-ray heated,
therefore, the incident spectrum is probably rather flat (consistent
with thermal bremsstrahlung) and the clouds must be thick ($>10^{22}$
cm$^{-2}$). But the mechanism is energetically unlikely based on ratios of
the observed X-ray luminosities to S(1) emission line luminosities, 
especially for
Abell 2597.

Mechanisms with physical processes similar to those which accompany  
X-ray heating, but with slightly
different sources of energy, have been postulated as energy sources
for the optical filaments. Conduction (Sparks \etal  1989) and irradiation
by mixing layers (Begelman \& Fabian 1990) are models which are not as 
easily tested by the data in hand, although the lack of coronal line
emission (Donahue \& Stocke 1994; Yan \& Cohen 1995) 
stongly constrains the parameters of mixing layer models, 
which rely on local
X-ray and UV-emitting gas to heat the filaments. 

All three models, conduction, mixing layers, and direct X-ray heating, tap into
the thermal energy of the ICM, but only conduction can draw
from the vast thermal reservoir not local to the filaments. 
Determining whether conduction or irradiation by mixing layers could 
generate sufficient luminosity to explain the observations depends on
assumptions about filament geometry, the electron density of the ICM,
and, to some extent, the magnetic field strength and structure. For 
example, a conduction model 
with modest assumptions about electron densities of $n_e\sim0.5$
cm$^{-3}$, $T_e\sim 10^7$K, filament surface area that is three times 
the projected area, and a conduction flux which is 1\% the classic
saturated flux, predicts there is sufficient energy in these systems 
to power the optical and infrared line luminosity of the filaments. The line
ratios of H$_2$ to H$\alpha$ have not been computed for such models, but
are likely to be similar to those we derive here for direct irradiation
of thick columns of hydrogen by the X-ray emitting ICM. Therefore we
cannot rule out conduction or mixing layers 
as a possible energy source for the molecular
emission-line filaments.

\subsection {Shocks}
 
The filamentary morphology of the emission is suggestive of shock heating,
but not proof. The observed optical line emission from 
the central 2" of A2597 is inconsistent
with shocks being the primary source of the energy (Voit \& Donahue
1997), but the molecular emission could arise from a separate 
process. In the next few paragraphs, we will examine the 
implications of our observations in the context 
of shock models (Hollenbach \& McKee 1989).

The intrinsic surface brightnesses predicted by the shock
models suggest that the
preshock densities are between $10^3$ and $10^4$ cm$^{-3}$, and
possibly up to $10^5$ cm$^{-3}$ for the brightest H$\alpha$ systems. 
If our line of sight passes  through multiple shock fronts, the observed 
surface brightness exceeds the true surface brightness of the 
working surface of any given shock, and the true preshock 
densities could be somewhat lower.

\begin{figure}
\plotone{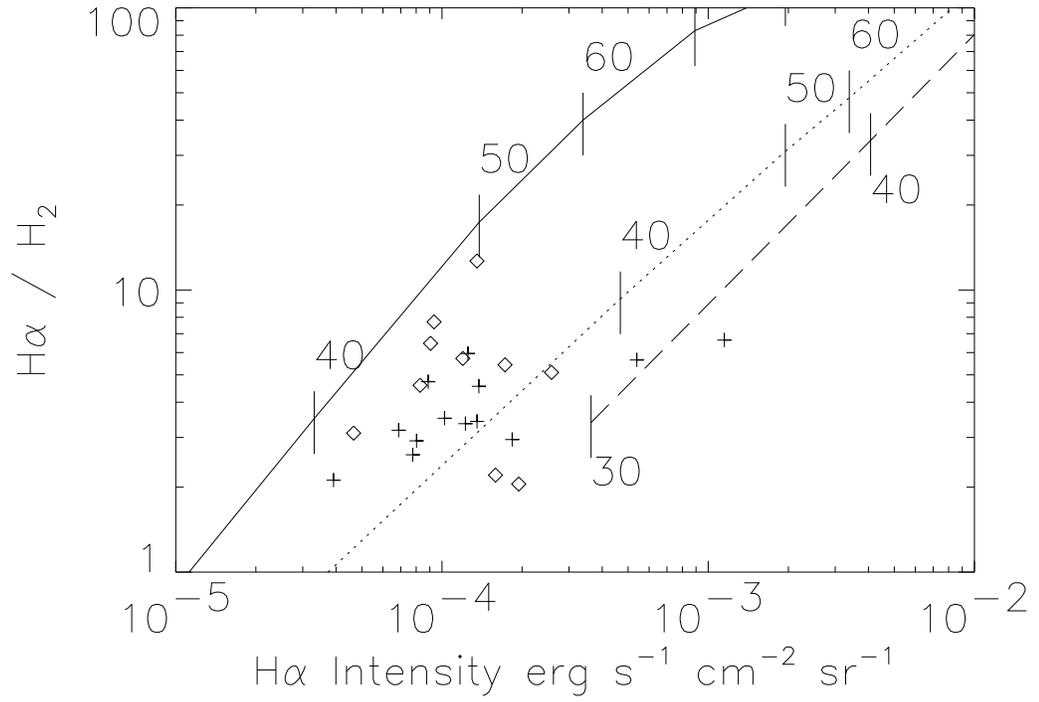}
\caption[]{Ratio of H$\alpha$ and H$_2$ emission line flux compared 
to the estimated H$\alpha$ surface brightness of the knots in
Abell 2597 (diamonds) and PKS0745-191 (crosses). The lines plot
the line ratios and surface brightness predicted by shock models
(Hollenbach \& McKee 1989)
with the $V_{shock}= 40-60$ km s$^{-1}$ labelled. The solid line
is for a preshock density of $n_0=10^3$ cm$^{-3}$ and the dashed
line is for $n_0=10^4$ cm$^{-3}$.\label{shocks}}
\end{figure}

Fast shocks produce H$\alpha$/H$_2$ line ratios that are significantly 
higher than observed, as fast shocks dissociate H$_2$ 
and have thick ionization regions. The observed  H$\alpha$/H$_2$ 
ratios span $1.5 -10$ in A2597 and $2-7$ in PKS0745-191 (the ratios
in PKS0745-191 are a factor of two higher if H$\alpha$ 
is significantly attenuated by dust). 
For preshock densities of $10^3 - 10^4$ cm$^{-3}$, the 
observed ratios constrain the shock 
velocities to be $<50$ and $<40$ km s$^{-1}$ respectively 
(Figure~\ref{shocks}). If the 
H$\alpha$ production is supplemented by stellar photoionization or any
other possible source, the 
H$\alpha$/H2 ratio from a putative shock is even lower, and thus the 
velocities of the shocks must be lower.

If low-velocity shocks are producing the H$_2$ emission, the amount of
matter being processed by these shocks is enormous, on the order
of 10,000 ­- 100,000 solar masses of shocked gas per year. Also, the
turbulent velocities must be significantly greater than the shock
velocities in order for such slow shocks to be consistent with the
velocity widths of the infrared and optical emission lines 
($v_{FWHM} \sim 300-500$ km s$^{-1}$.)
However, models of radio lobes and the shocks propagating into a
uniform medium suggest that in some cases, one might
predict extremely slow shocks if the preshock gas is very dense, 
while the plume of material behind
the shock may emit substantially broader optical emission lines
(Koekemoer, private communication.)

\subsection {UV Fluorescence and Star Formation}
\subsubsection{Star Formation, H$\alpha$, and Far Infrared Emission}

If the H$\alpha$ in the central 2" in 
all three galaxies is generated by star formation, we can infer a 
lower limit for steady star formation (a lower limit
because the H$\alpha$ could be 
absorbed by dust) for the three targets of 2-5 h$^{-2}$ M$_\odot$ yr$^{-1}$,
scaling tabulated models from Gehrz, Sramek, \& Weedman (1983). We explored
a standard model with initial mass function slopes of $\alpha=3.5$ and 
initial stellar masses between 6-25 solar masses.  (We note that these star
formation rates fall 1-2 orders of magnitude short of that required to explain
the fate of the mass inferred from X-ray observations to cool from the intracluster
media.) Much of the radiation
generated by star formation in dusty environments 
is re-emitted in the far infrared. 
The bolometric (infrared) luminosities
predicted from these models, as required to generated the observed
H$\alpha$ luminosities, are $8-23 \times 10^{43} \lum$, where
the predicted luminosity of NGC1275 is $8 \times 10^{43} h^{-2} \lum$, 
A2597 is $1 \times 10^{44} h^{-2} \lum$, and PKS0745-191 is $2 \times 10^{44} 
h^{-2} \lum$, 
consistent with and not too far below the 2$\sigma$ 
IRAS upper limits for 4' diameter apertures centered on 
Abell 2597 and PKS0745 (Wise \etal 1983). 
NGC1275 is a FIR source (Moshir \etal  1990)
with $\nu L_\nu$ at 60 microns of $1.2\times10^{44} h^{-2} \lum$. ($\nu L_\nu$
of a galaxy at 60 microns is within a factor of several of the total 
FIR luminosity of the galaxy.) FIR luminosities $\nu L_\nu$ of A2597 and
PKS0745-191 have non-restrictive upper limits of $<2\times10^{44} h^{-2} \lum$
and $<7\times10^{44} h^{-2} \lum$ respectively (Wise \etal  1993). 
The far-infrared observations of these galaxies are
consistent with the conjecture that all of the H$\alpha$ may be produced
as a result of star formation. We will next test whether this assumption is
consistent with the near-infrared observations.

\subsubsection{UV Fluorescence, H$\alpha$, and H$_2$ Emission}

If the emission is powered by UV fluorescence, the vibrational
excitation states are maintained indirectly by absorption of ultraviolet
starlight in the Lyman and Werner band systems, followed by fluorescence
(e.g. Black \& van Dishoeck 1987).
We can infer the intrinsic UV intensity by extrapolating from 
the ionizing flux required to produce the H$\alpha$ photons to the 
UV flux that would be available to excite the vibrational states of H$_2$. 
If the H$\alpha$ emission arises from photoionization, one can 
estimate the ionizing flux  ($\lambda<912$\AA) 
required to produce it. If these same photoionizing sources also produce
the $1100-912$\AA~ photons, then with some assumptions about
the shape of the incident spectrum, one can estimate the local intensity
of $1100-912$\AA~ flux, which excites the vibrational transitions of
H$_2$ gas.

For UV photoionization, 
0.45 H$\alpha$ photons emerge for every ionizing photon absorbed, 
and $\sim0.2\%$ of the incident UV energy between 1130 and 912 \AA~ 
emerges in a single strong 
H$_2$ line (this efficiency ranges between $0.06\%-0.3\%$; BvD87). Then 
a typical power-law spectrum for an AGN produces an
H$\alpha$/H$_2$ ratio of $\sim80$ ($60-300$), 
assuming $s=0.86$ longward of
912 \AA~ and $s=1.8$ shortward of 912\AA, where $f_\nu \propto \nu^{-s}$
(e.g. Shull \etal 1999).
An ionizing/UV background dominated by stellar sources can be
approximated by inserting a break at 912 \AA. A spectrum with a break factor 
of $\sim100b_{100}$, 
where $b_{100}=1$, typical
of Galactic UV sources, generates significantly lower 
H$\alpha$/H$_2$ ratios of $\sim3-0.5 / b_{100}$, closer to what we observe.

We compare the observed UV flux to that required to 
produce the H$_2$ emission, or the H$_2$ and H$\alpha$ emission, 
and derive the implied extinction.
IUE archival spectra show UV continua   
between $11-13.6$ eV for all of three targets. 
The UV continuum emission from these sources is very
compact and essentially unresolved by IUE, and implying an extent less 
than 3". In A2597 and PKS0745-191, the UV flux at a rest wavelength of 
about 1100 \AA~ is $F_\lambda \sim 10^{-15} \flux$~\AA$^{-1}$, thus the 
luminosity per unit wavelength is 
$\sim 10^{39} \lum$~\AA$^{-1}$. (For 
NGC1275, the flux of $10^{-14} \flux$~\AA$^{-1}$ translates to a specific
luminosity of  
few times $10^{39} \lum$~\AA$^{-1}$.)

Using the same spectral shape from Shull \etal (1999) and a break factor
of $100$, the observed fluxes implied by the H$\alpha$ luminosities from
the central $r=2"$ aperture are $\sim5 \times 10^{-14} \flux$~\AA$^{-1}$. 
These fluxes are consistent with the IUE fluxes if $A_V \sim 1.1-1.2$ 
(Seaton 1979). 

We can also compare the H$_2$ surface brightness to the lower limit of
the UV surface brightness, as estimated from IUE observations. 
BvD87 define the dimensionless quantity $I_{UV}$ in their models  as 
$I_{UV}=I / 4.76\times 10^{-5}$ 
erg s$^{-1}$ cm$^{-2}$ sr$^{-1}$ between $912-1130$\AA. The observed 
$I_{UV}$ in a 3" by 3" aperture by this definition is $\sim30$ for 
A2597 and PKS0745-191, including a $(1+z)^4$ surface brightness 
correction. $I_{S(1)}$ ranges from $10^{-4}$ to $10^{-5}$
erg s$^{-1}$ cm$^{-2}$ sr$^{-1}$ in A2597 (values from Table~\ref{a2597_phot_table}), twice those values in PKS0745-191 
(Table~\ref{pks_phot_table}).
In fluorescent excitation models, a single bright line of H$_2$ represents
about 0.015 of the total amount of infrared line emission, so that
$I_{IR} \sim I_{S(1)}/0.015 \sim 10^{-2}-10^{-3}$ erg s$^{-1}$ cm$^{-2}$ sr$^{-1}$. The $I_{UV}$
required to produce such a surface brightness is $100-1000$ , with
a minimum total hydrogen density of 1000-3000 cm$^{-3}$ (BvD87, Figure
5).
A dust screen with an 
$A_V$ of only $\sim 0.5-1.1$ magnitudes from Galactic dust, corresponding
to a UV extinction factor of $\sim 3-40$ (Seaton, 1979), would be 
required to reconcile UV continuum observations with that required
to produce the H$_2$. The UV extinction implied is 
consistent with the approximate 
extinction derived from the Balmer line ratios
from the nucleus of A2597 (Voit \& Donahue 1997) and from our very 
rough estimates of the H$\alpha$/P$\alpha$ ratios. Therefore, based on
current limits on the intrinsic absorption, the observed
UV flux from these objects can not  
rule out UV irradiation by stars on energetic grounds.

To summarize, based on the H$\alpha$/H$_2$ line ratios, 
we can rule out a power-law source of UV photons heating the gas in
A2597 and PKS0745-191 but we cannot rule out 
stellar UV sources. The line ratios and the energetics are 
consistent with the possibility that the stars are heating both the
ionized gas and the vibrationally-excited molecular gas.

If UV radiation is the major heating source, the fluorescence models, 
normalized by the amount of stellar UV photons needed to produce the 
H$\alpha$ emission lines,
predict that there should be a significant number of $v=2$ vibrational
emission lines. Such
lines, like 2-1S(3) (2.07 $\mu$m), 2-1S(1) (2.25 $\mu$m) do not appear in the 
spectra of Falcke \etal  
(1998), but the most significant test comes from the
spectral region between 1-2 microns, which should be relatively rich in 
H$_2$ emission lines if UV fluorescence were important (BvD87). This line
region is often obscured by atmospheric line emission, and would be easiest
to observe from space (but some windows through to the ground exist for
clusters at specific redshifts).

Furthermore, if UV fluorescence is the major heating source, a significant
fraction of the UV will emerge as far-infrared radiation from dust. Such a
scenario predicts that both Abell 2597 and PKS0745-191 should 
be moderately strong FIR emitters
of $\sim 10^{44} h^{-2} \lum$, like NGC1275. SIRTF observations of these 
sources would quickly confirm or disprove this hypothesis.

A significant consequence of UV fluorescence as the most important heating
source compared to thermal or shock heated gas is the larger amount of
molecular hydrogen required to be present to produce the emission.
The difference in mass of molecular hydrogen if the gas is
excited by UV and a simple thermal model with $T=2000$ K, 
which would apply in the case of shocks or X-ray heated gas,
is a factor of a million. 
With model 14 of BvD87, which is
of an $n_H=3000$ cm$^{-3}$ cloud with an incident UV flux of 
$I_{UV}=1000$,  $L_{40} = L_{1-0~S(1)}/10^{40} \lum$ can be produced
with $1.1 \times 10^{9} L_{40}$ M$_\odot$ of H$_2$ , 
while the same amount
of luminosity can be produced from only $4.1 \times 10^{3} L_{40}$ M$_\odot$ of 
H$_2$ heated to 2000 K (Model S2, BvD87). For the 1-0~S(1) luminosities
from our sources, the H$_2$ masses range from a few billion solar
masses for the UV irradiation models to ten thousand solar masses for
the thermal models ($\propto h^{-2}$).

For PKS0745-191, $5\times10^9 h^{-2}$ solar masses of H$_2$ is
implied, which is very close to the 
published limits of H$_2$ masses from 
CO observations assuming the standard CO/H$_2$
conversions by O'Dea \etal (1994). We note the CO/H$_2$
conversion is large and likely not accurate in these environments. 
NGC~1275 has a uncertain H$_2$ mass, based on CO observations 
with the standard conversion, of $3\times10^9 h^{-2}$ M$_\odot$
(Lazareff \etal 1989), very
similar to the $3\times10^{9} h^{-2}$ M$_\odot$ predicted by the
H$_2$ flux and UV excitation models. This correspondence cannot
be taken too seriously because of the conversion uncertainty. 
Abell~2597 does not have published observations of CO emission.  

\section{Conclusions \label{conclusions}}

We have discovered {\em extended} kiloparsec-scale 
vibrationally-excited molecular hydrogen in the 
cores of galaxies thought to be in the 
centers of 
cooling flows in the clusters Abell 2597, PKS0745-191, and Perseus/NGC1275. 
The molecular gas was already known to be present in the nuclei, from ground-based
spectroscopy, but its structure was unknown. We confirm the extended structure
of molecular hydrogen seen in NGC1275 by Krabbe \etal (2000) from the ground. 
The vibrationally-excited molecular emission appears morphologically very 
similar to the ionized nebular gas in Abell 2597 and PKS0745-191. We have also
discovered dust lanes which are optically thick to 1.6$\mu$m emission.
These dust lanes do not trace the nebular gas, but are also in the 
central few kpc. NGC1275 has prominent dust lanes at 1.6$\mu$m. We
have shown that the dust reddening in NGC1275 is consistent with that of 
Galactic dust. We have also, as a byproduct of our observations, obtained 
H-band (1.6$\mu$m) photometry of NGC1275 globular clusters. We have
been able to map the H$\alpha$/H$_2$ ratios in different positions 
across A2597 and PKS0745-191.

We examine the possible sources of the vibrationally-excited molecular gas, 
which cannot have
condensed from a cooling flow. We also investigate the source of heat
for the vibrationally-excited molecular gas. 
 The source of heat in NGC1275 is likely the
central AGN -- we cannot discriminate between the possible processes
there. For the extended emission in 
A2597 and PKS0745-191, we can rule out AGN photoionization and fast
shocks because the H$\alpha$ / H$_2$ ratios are 
too low. We can rule out X-ray heating for
A2597 because of insufficient  
X-ray intensity local to the filaments. 
Extremely slow shocks ($<40$ km s$^{-1}$) 
produce significantly higher H$_2$/H$\alpha$ ratios than do fast
shocks, consistent with what we observe in A2597 and PKS0745-191, 
but not very efficiently. In order to produce a constant 
observed luminosity, tens of thousands of solar masses of gas 
must be processed by the shocks in each source each year. We conclude that 
the radio sources are not injecting significant amounts of energy 
by strong shocks into the interstellar and intracluster
medium -- at best, the emission line and molecular gas may be pushed
around by the radio sources.
 
UV irradiation by very hot stars, implied by a star formation rate of only
a few solar masses per year, 
is a possibility yet consistent with both 
the H$\alpha$ / H$_2$ line ratios and UV continuum observations by IUE. 
$1-2$ micron spectroscopy would definitively test
this hypothesis because UV-fluorescence produces significant amounts of
$v=2$ vibrational lines from excited molecular hydrogen. Also, 
if UV irradiation is the dominant mechanism, the 
far-infrared ($\sim60 \mu$m) luminosities should be not far below
the IRAS upper limits inferred for A2597 and PKS0745-191, $\sim10^{44} h^{-2}
\lum$, easily detectable by SIRTF observations.  
Thermally-heated molecular gas very efficiently 
radiates molecular hydrogen lines so in the case of PKS0745-191, 
the ambient X-ray gas might heat a small amount of molecular hydrogen. We
cannot rule out conduction and turbulent mixing layers as plausible
heat sources, although energetically, direct X-ray heating and turbulent
mixing layers are unlikely. We conclude that UV irradiation  
is the most viable heat source for
both PKS0745-191 and A2597. If UV fluorescence is the dominant physical
mechanism, $\sim3\times10^{9}$ M$_\odot$ 
of vibrationally-excited H$_2$ exist in these 
cluster galaxies, while if the gas is thermally excited by X-rays or
conduction, only $\sim10,000$ M$_\odot$ in each galaxy are implied. 

These observations provide some insight into the physical processes that
may be occurring in the cores of clusters with cooling flows. 
Vibrationally-excited H$_2$ and dust lanes appear 
to exist in the same cooling flow clusters with optical emission-line nebulae. 
The H$_2$ emission is too bright to be produced directly by the cooling
flow. The amount of H$_2$ implied to be present is at least a factor of
100 too small to account for a repository of molecular gas that has 
accumulated over a Hubble time of constant cooling, even if the emission
is powered by the UV radiation from star formation. The star formation rate
inferred from the H$\alpha$ and H$_2$ emission and an assumption of UV-powering is also too small to 
account for cooling rates of $>100$ M$_\odot$ yr$^{-1}$. However, these 
observations also show that the radio sources in these clusters do not
impart a significant amount of shock energy to the emission line gas, and
therefore cannot be a significant source of heating to counterbalance the
cooling in the centers of cooling flows. 

\acknowledgements
This work was supported by the NASA HST grant GO-7457. 
PRM is supported by the NASA Astrophysical Theory Program under grant
NAG5-4061 and by NSF under grant AST-9900871.This work is based partly on 
observations made with the NASA/ESA Hubble Space 
Telescope obtained from the data archive at the Space Telescope Science Institute. 
Some of the data presented in this paper were obtained from the Multimission 
Archive at the Space Telescope
Science Institute (MAST). Support for MAST for non-HST data is provided by 
the NASA Office of Space Science via grant NAG5-7584 and by other grants 
and contracts. STScI is operated by the 
Association of Universities for Research in Astronomy, Inc. 
under NASA contract NAS 5-26555. This research has also made use of data
obtained from the High Energy Archive Research Center (HEASARC), provided
by NASA's Goddard Space Flight Center. 
We acknowledge extensive discussions and data reduction advice 
from David Zurek. We thank Eddie Bergeron, John Biretta, 
Mark Dickinson, Paul Goudfrooij, 
Roeland van der Marel, Cathy Imhoff and Sylvia Baggett for very useful 
discussions. We are grateful to the anonymous referee for his or her comments. 
This research has made use of the NASA/IPAC Extragalactic Database (NED), which
is operated by the Jet Propulsion Laboratory, California Institute of Technology,
under contract with the National Aeronautics and Space Administration.

\appendix
\section{Globular Clusters and Dust in NGC1275}

The 1.6 micron image obtained in our program, co-added to a similar image
taken for a parallel observing program, has yielded near infrared photometry for the
NGC1275 star clusters first detected by Holtzmann \etal 1992. We have
extracted WFPC2 photometry for the same clusters in order to enable 2-color
studies with the ``H''-band photometry. Our photometry procedure and results
are reported here. We compare the colors obtained for these clusters with
those predicted for a young stellar population with a single burst of
star formation as a function of time since the burst, 
and we compare our results to those
of optical HST photometry of the same star clusters 
(Carlson \etal  1998, C98 henceforth).

\subsection{Globular Cluster Photometry}

	The large number of centrally distributed globular clusters in NGC1275 
allows an estimate of individual globular cluster colors by comparing photometry
with WFPC2 and NICMOS at B, R, and H. We use both our F160W data
and archival data (Regan \& Mulchaey 1998, Program 7330) 
to create a summed H-band image. 
The F450W and F702W  archival images from Program 6228 (Carlson \etal 1998; 
Holtzman \etal 1996, 1992) were deep B and R images of the globular clusters 
in NGC1275, with exposure times of 4900 and 4500 seconds, respectively.

\begin{figure}
\plotone{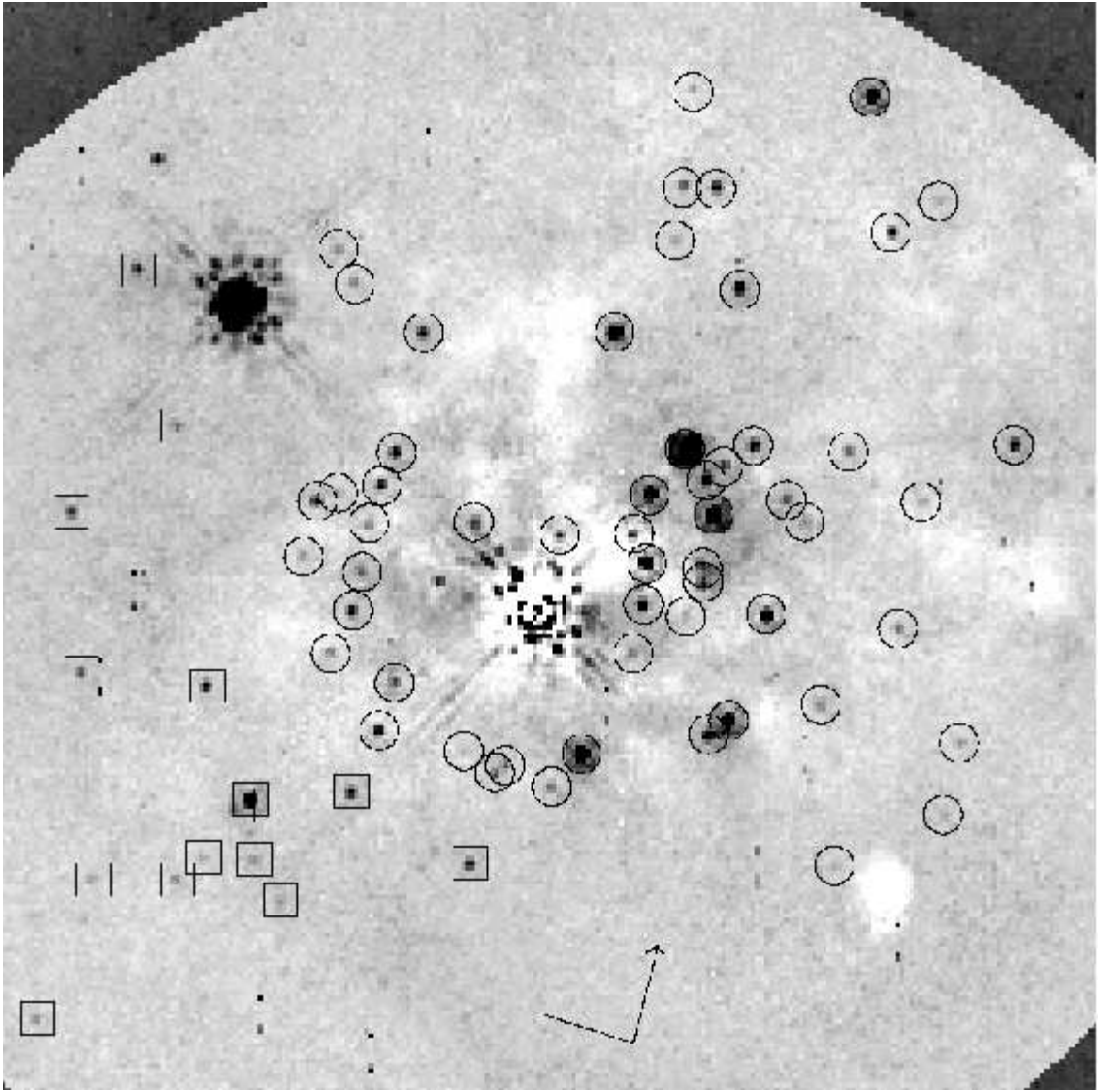}
\caption[]{Star clusters in NGC1275.  An ellipsoidal model has been 
subtracted from the 1.6-$\mu$m HST data. The white regions outline the 
dust lanes; the dark spots are star clusters. The nucleus and a foreground
star appear as prominent point sources with diffraction spikes and Airy rings in
the center and the upper left of the image. The circular white absorption
feature in the lower right hand corner is the coronograph. 
The edges of the ellipsoidal model are visible in the corners of the image. 
The boxed clusters lie in the same lines of site as the regions which
appear to be free of dust lanes. The circled objects indicate other clusters
which most likely lie in the
same lines of site as the dust lanes. North is marked with an arrow. 
\label{glob_picture}}
\end{figure}

In Figure~\ref{glob_picture}, 
we present the 640-second exposure 1.6 $\mu$m 
image with the best fit elliptical 
model subtracted and with globular cluster candidates marked. Compact
clusters which were apparent in the NICMOS image and both the F450W and F702W 
images were chosen for comparison.  Aperture photometry was performed using
DAOPHOT, with the aperture size chosen to be the radius where the source point
spread function (PSF)   
fades into the noise. This corresponds to 4 pixels for NICMOS (0.30") and 3 
pixels for the PC on WFPC2 (0.14"). Sky values were determined by estimating the mode of 
background in an annulus from 4-7 pixels for WFPC2 and from 5-8 pixels in 
NICMOS. We correct the fluxes to $r=0.5\arcsec$ apertures by  
generating the PSF for each filter with Tiny TIM, matching the FWHM of
the convolution of the model PSF with a 2D Gaussian to the FWHM of the 
average globular cluster, and estimating the appropriate flux correction.

The photometric zeropoints and color transformations were derived using 
SYNPHOT, allowing us to calculate the total magnitude and color of each 
globular cluster for Johnson B- and R- photometry and for HST H-photometry.  
The zeropoint for the WFPC2 data (VEGAMAG) includes an aperture 
correction from $r=0.5"$ to ``infinite'' aperture (0.10 mag). The NICMOS data
zeropoints, on the other hand, are given for an $r=0.5"$ aperture. 
To correct to 
``infinite'' aperture, we multiplied the total count rate by 1.15, as 
recommended by the NICMOS Photometry section on the \anchor{http://www.stsci.edu/instruments/nicmos/topnicmos.html}{NICMOS 
instrument webpage}. The
zeropoints for each filter are listed in Table~\ref{phot_system}.

To derive the best Johnson color transformations, 
the spectral type of the objects must be known in advance. Brodie et al. (1998)
compare the spectra of 5 globular clusters in NGC1275 with 
standard stellar spectra covering a range of spectral types and luminosity 
classes. The best match to these spectra, early type A dwarfs and class III 
giants, was used as the model for transforming the HST B and R filters to the 
Johnson system. The resulting color corrections are negligible, 0.007 and 0.001 mag for each 
filter respectively. The HST VEGAMAG 
system was used for the cluster H-band photometry. 
For reference, the color term required to correct the HST system to 
the KPNO H-system  is +0.034 mag. The corrections used for our 
photometric calibration 
are summarized in Table~\ref{phot_system}.

\begin{table*}
\caption{Photometry Corrections for Star Clusters in NGC1275 \label{phot_system}}
{\begin{tabular}{lrrr}
\tableline
\multicolumn{1}{l}{Magnitude System} &
\multicolumn{1}{c}{Vega} &
\multicolumn{1}{c}{Aperture} &
\multicolumn{1}{c}{Color}\\
\multicolumn{1}{l}{(magnitudes) } &
\multicolumn{1}{c}{Zeropoint} &
\multicolumn{1}{c}{Correction} &
\multicolumn{1}{c}{Correction} \\

\tableline \tableline
Johnson B & $21.99 \pm0.02$  &  -0.17  & -0.01 \\	
Johnson R & $22.43 \pm0.02 $ &  -0.21  & -0.00 \\	
    HST H & $21.67 \pm0.05$  &  -0.08  &       \\	
\tableline
\end{tabular}}
\end{table*}

C98 presented an analysis of the compact star clusters in NGC1275
using deep WFPC2 images at B- and R-bands, the same data we retrieved from
the archive. They present B-R color vs B magnitude plots for all clusters
in the PC and the WFs and separately for clusters in the relatively dust-free 
southwest portion of the PC. This analysis revealed a bimodal color 
distribution, with a blue population having (B-R)$_0$=0.4 and a red population 
with (B-R)$_0$=1.3. The red objects are members of an old globular cluster 
system, while the blue clusters are a much younger population. The inferred age
of the blue star clusters was derived using Bruzual-Charlot (1993) models and the 
measured B-R colors, indicating a range from 10$^7$ to 10$^9$ yr.

\begin{figure}
\plotone{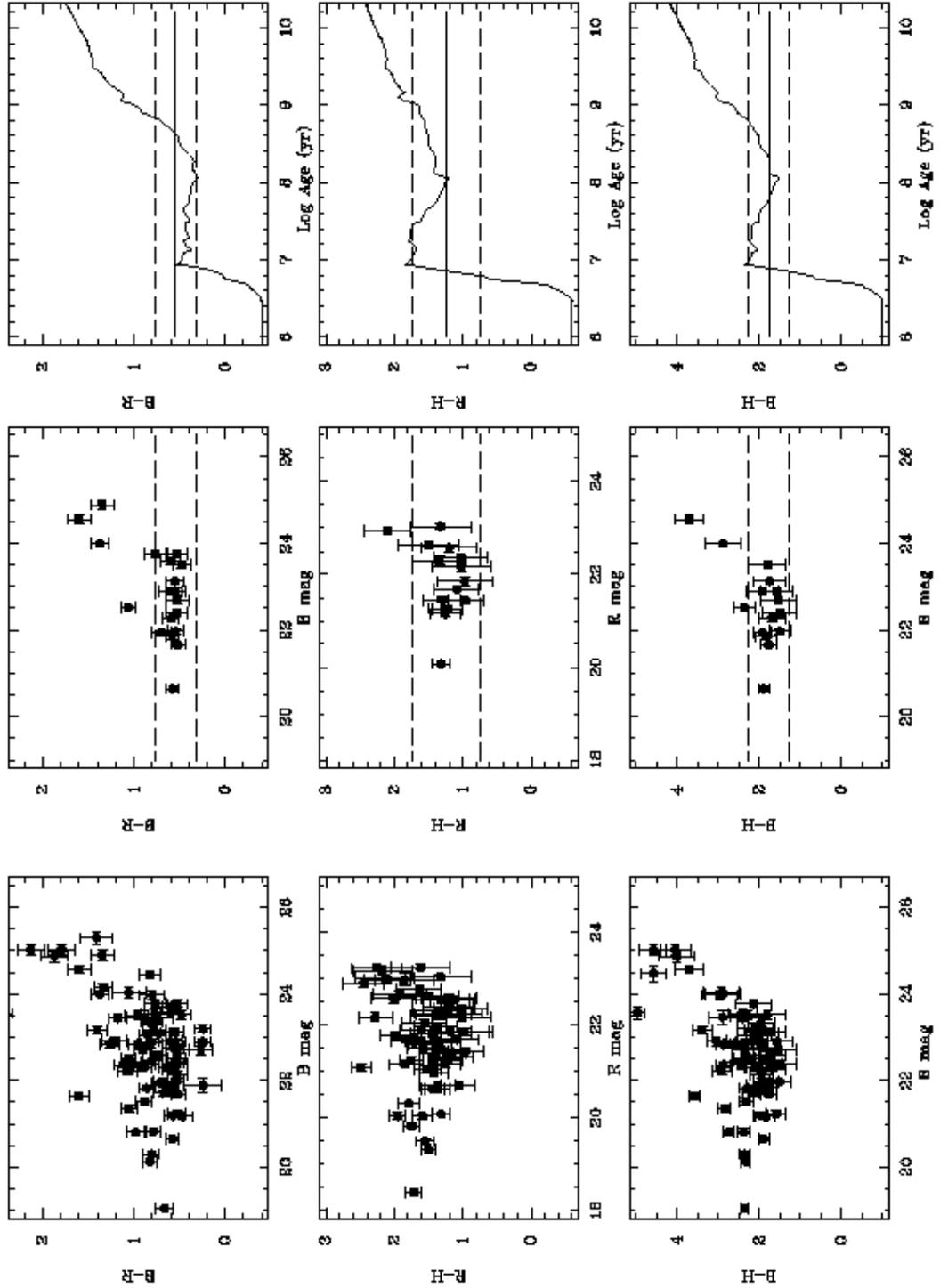}


\caption[]{Star Cluster Photometry and Age Determinations. 
Photometry points with $1\sigma$ error bars. 
Dashed lines in plots in columns 1 and 2 
mark the $3\sigma$ dispersion of the colors  
for the dust-free sample of blue clusters.
(Column 1) Magnitude-color plots for all clusters. 
(Column 2) Magnitude-color plots for the clusters in the non-dusty regions. 
(Column 3) $3\sigma$ color limits (dashed lines) and 
average colors (solid straight line) 
overlaid on single-burst 
Bruzual-Charlot model (solid lines) of color vs. age. This plot shows that the 
cluster colors are consistent with ages of $10^7-10^9$ years.\label{glob_plot}}
\notetoeditor{Figure 11 is in Landscape orientation.}
\end{figure}

Because the extinction is so much less at H-band than at B or R, 
the B-H colors could produce an age estimate less dependent on 
assumptions about dust. 
In Figure~\ref{glob_plot}a, we present 
the photometry of all objects detected at 1.6 $\mu$m by plotting color against 
apparent magnitude, assuming Galactic extinction to be 0.71, 0.40, and 0.10 
magnitudes at B, R, 
and H respectively. 
Following the methods of C98, we tested whether the 
observed scatter in color was a result of internal extinction. Therefore, for 
this analysis, we chose clusters from the isophotal residual image which were 
clearly separate from the dusty regions. These clusters are indicated in 
Figure~\ref{glob_picture} and are plotted in Figure~\ref{glob_plot}b.

We find that the majority of our selected objects belong to the blue population 
described by C98, 
although we do detect a few of the red clusters, which are intrinsically 
fainter. The dominance of blue population objects is 
due to the shallower depth of the H-band data, thus limiting our 
sample to the brighter (and therefore bluer) clusters in the field. By limiting 
the photometry to dust-free regions of the image, the scatter in the observed 
(B-R) colors decreases from 0.23 mag to 0.08 mag. This scatter is consistent with 
the photometric errors, supporting the idea that varying amounts of
dust extinction, rather 
than differences in age, are responsible for the dispersion in the measured 
colors of each population. 

The blue clusters have a typical (B-R) color of 0.5 mag and (B-H) color of 1.7 
mag. Figure~\ref{glob_plot}, third column,  
shows the Bruzual-Charlot (1993) color evolution of a single-burst 
population, assuming a Salpeter IMF from 0.1 to 125 M$_o$ and a total mass of 
1 M$_o$. Dashed lines indicate the $3\sigma$ dispersion of the blue cluster 
photometry, allowing us to infer ages from 10$^7$ to 10$^9$ yr. 
The effect of eliminating 
internal extinction by performing the photometry in the infrared has little
effect in constraining the age of these clusters, since the B-R and B-H
colors are insensitive within this age range.

Therefore, we confirm the age estimates of C98 by using
optical-near infrared colors of the star clusters. We have also demonstrated,  
by selecting a sub-sample of stellar clusters from the dust-free regions
of the infrared image, that the scatter in B-R color is consistent 
with a scatter in the intrinsic dust absorption and not in cluster 
properties such as age.

\newpage

\clearpage

\end{document}